\documentclass{aa}
\usepackage{graphicx}
\usepackage{natbib}
\usepackage{txfonts}
\usepackage{epstopdf}
\bibpunct{(}{)}{;}{a}{}{,}    % to follow the A&A style

\def\apjs{ApJS}
        
\def\jgr{J. Geophys. Res.}
          
\def\grl{Geophys. Res. Lett.}

\def\aap{Astron. Astrophys.}

\def\apj{Astrophys. J.}

\def\solphys{Sol. Phys.}
     
\def\nat{Nature}

\def\mnras{Mon. Not. R. Astron. Soc.}

\begin{document}
\title{The variability of Sun-like stars: reproducing observed photometric trends}
\author{A.I. Shapiro \inst{1} \and S.K. Solanki \inst{2,3} \and N.A. Krivova \inst{2} \and W.K. Schmutz \inst{1}  \and W.T. Ball \inst{4} \and R. Knaack \inst{5} \and  E.V. Rozanov \inst{1,6}  \and \\ Y.C. Unruh \inst{4}}
\offprints{A.I. Shapiro}

\institute{Physikalisch-Meteorologishes Observatorium Davos, World Radiation centre, 7260 Davos Dorf, Switzerland\\
\email{alexander.shapiro@pmodwrc.ch}
\and Max-Planck-Institut fur Sonnensystemforschung, Goettingen, Germany 
\and School of Space Research, Kyung Hee University, Yongin, Gyeonggi 446-701, Korea
\and Blackett Laboratory, Imperial College London SW7 2AZ, United Kingdom
\and Hochschule fur Angewandte Wissenschaften, Zurich, Switzerland
\and Institute for Atmospheric and Climate science, ETH, Zurich, Switzerland\\}

\date{Received ; accepted }

\abstract
% context heading  (optional}
{The Sun and stars with low magnetic activity levels, become photometrically brighter when their activity increases.  Magnetically more active stars display the opposite behaviour and get fainter when their activity increases. }
% aims heading (mandatory)
{We reproduce the observed photometric trends in stellar variations with a model that {  treats stars as hypothetical Suns with coverage by magnetic features  different from that of the Sun}.}
% methods heading (mandatory)
{The presented model attributes the variability of stellar spectra to the imbalance between the contributions from different components of the solar atmosphere, such as dark starspots and bright faculae. A stellar spectrum is calculated from spectra of the
individual components, by weighting them with corresponding disc area coverages. The latter are obtained by extrapolating the solar  dependences of spot and facular disc area coverages on chromospheric activity to stars with different levels of mean chromospheric activity. } 
% results heading (mandatory)
{We have found that the contribution by starspots to the variability increases faster with chromospheric activity than the facular contribution. This causes the transition from faculae-dominated variability and direct activity--brightness correlation to spot-dominated variability and  inverse activity--brightness correlation with increasing chromospheric activity level. We have shown that the regime of the variability also depends on the angle between the stellar rotation axis and the line-of-sight and on the latitudinal distribution of active regions on the stellar surface. Our model can be used as a tool to extrapolate the observed photometric variability of the Sun to  Sun-like stars at different activity levels, which makes possible the direct comparison between  solar and stellar irradiance data.}
{}


\titlerunning{The variability of Sun-like stars}
\maketitle

\section{Introduction}\label{sect:intro}
The activity cycles in lower main sequence stars were discovered at the Mount Wilson Observatory by recording the emission in the cores of H and K lines of Ca II \citep{wilson1978}. The time-variable part of this emission forms mainly in the bright chromospheric areas, heated by the magnetic field, and therefore is usually linked with  stellar magnetic activity, which is also often referred to as chromospheric activity \citep[see e.g.][]{HK,Hall_LR}. The Mount Wilson program revealed that 60\% of the lower main sequence stars exhibit periodic variations in activity reminiscent of the solar 11-year cycle,  25\% show fluctuations of activity without a clear periodicity (these stars are generally young and magnetically active), and 15\% are the activity-flat stars whose activity is constant with time  \citep{baliunasetal1998}. This distribution was confirmed  with the complementary synoptic Solar-Stellar Spectrograph program at the Lowell Observatory, which was more focused on nearly Sun-like stars \citep[{  defined by \cite{lockwoodetal2007} as stars on or near main sequence and colour index $0.42 \le (B-V) \le 1.4$,}][]{halletal2007b, Hall_LR}.

Stellar photometric variations have been more difficult to detect, which, in fact,  is also true in the case of the Sun. The solar S-index (which is a measure of the Ca II H and K emission)  varies by more than 10\% over the 11-year cycle \citep{radicketal1998}, while the corresponding variability of the Total Solar Irradiance (TSI, which is the spectrally integrated solar radiation at one Astronomical Unit from the Sun) is only about 0.1\% \citep{pmod_comp}.  The first synoptic observations in the Str{\"o}mgren filters {\it b} and {\it v} (centreed at 467 and 411 nm, respectively) of dwarf stars close to the solar spectral type   were not able to unambiguously confirm their variability \citep{Lowell_start}. 

A considerable improvement in the precision of ground-based stellar photometry made it possible to establish the variability of the young main-sequence stars in the Hyades \citep{radicketal1995}. Their photometric variations in the Str{\"o}mgren {\it b} and {\it y}  filters (centreed at 467 and 547 nm, respectively) appeared to be in antiphase with variations of  chromospheric activity \citep[see also][]{lockwoodetal2007}.

A significant milestone was achieved with the start of the synoptic observations of Sun-like stars at the Lowell observatory \citep{lockwoodetal1992}.  \citet{lockwoodetal1997} selected 41 program stars bracketing the Sun in temperature and mean chromospheric activity (34 of these stars were also included in the Mount Wilson survey)  and started their regular photometric monitoring in the Str{\"o}mgren filters {\it b} and {\it y}. Most of the stars in the program were deemed variable on both the stellar rotation and the cycle time scales 
\citep{lockwoodetal1997, radicketal1998, lockwoodetal2007}.  While the solar 11-year variability in the Str{\"o}mgren {\it b} and {\it y} filters is estimated to be around 0.0002 mag \citep{shapiroetal2013_stars}, the variability of several stars was found to exceed 0.01 mag. Hereafter the variability is calculated in terms of root mean squares (rms), which is more appropriate for the analysis of the rather large variety of stellar light curves than the amplitude \citep{radicketal1998}.

One of the most robust patterns of stellar variability established by the Lowell program  was a conspicuous division between less active \citep[and consequently older, see e.g.][]{soderblometal1991} and more active (younger) stars. The former usually demonstrate a direct correlation between brightness and activity on the activity cycle time scale (an increase in activity is accompanied by an increase in the photometric brightness). The latter, on the contrary, show an inverse correlation between brightness and activity, i.e. these two quantities vary in antiphase. The threshold between these two regimes occurs around an average chromospheric activity of ${\rm log R'_{ HK}} \approx -4.7$ \citep[see][for a definition of  ${\rm log R'_{ HK}} $]{radicketal1998}, which roughly corresponds an age of 2 Gyr age \citep{Hall_LR}. We note that the mean level of solar chromospheric activity is ${\rm log R'_{ HK}} \approx -4.895$ \citep{lockwoodetal2007} and the Sun's age is estimated to be around 4.5~--~5 Gyr \citep[see e.g.][and references therein]{dziembowskietal1999}. Thus, according to this classification the Sun belongs to the group of older stars with direct activity-brightness correlation.

This result  was later confirmed by \cite{halletal2009} based on 14 years of combined photometric and Ca II measurements of 28 Sun-like stars taken with Automatic Photometric Telescopes at Fairborn Observatory and the Solar-Stellar Spectrograph at Lowell Observatory; seven of these stars were also members of the \cite{lockwoodetal2007} sample; the sample included all the brightest solar analogs with low activity. \cite{halletal2009} considered the activity-brightness correlation in four-year moving boxcars and found that HD 140538 ($\psi$ Ser with a mean chromospheric activity of $ {\rm  log R'_{ HK}} =- 4.80$)  lies exactly at the threshold between the two variability regimes and exhibits both direct and inverse activity-brightness correlations on timescales of four years. They also found that besides the chromospheric activity there must also be other factors  which affect the regime of  variability. For example  HD 82885  ($ {\rm  log R'_{ HK}} =- 4.70$) showed the most robust inverse activity-brightness correlation among the entire \cite{halletal2009} sample, being just slightly more active than HD 140538.  Up to now there has been no good model to explain these observations.

The Lowell and Fairborn programs have also shown that stars with higher chromospheric activity have, in general, larger photometric variability.  \cite{lockwoodetal2007} and \cite{halletal2009} presented empirical linear dependences of variability (expressed via $\log {\rm   (rms(b+y)/2)  } $) on activity (as traced by $ {\rm  log R'_{ HK}} $). These regressions have the same slopes, but the \cite{halletal2009} regression is shifted by about 0.1 dex down in variability relative to the  \cite{lockwoodetal2007}  regression.

One of the crucial  issues in studies of the solar-stellar connection is understanding whether solar variability obeys the empirical patterns outlined above. Presently there is no clear answer to this question. While the regime of the TSI variability over the 11-year cycle is well-established -- it varies in phase with solar activity \citep[see e.g.][and references therein]{pmod_comp, balletal2012, MPS_AA}, 
the regime of  the solar variability in the visible part of the spectrum is still under debate. The Sun is located so close to the threshold  between direct and inverse correlations of visible irradiance (e.g. as measured in the Str{\"o}mgren {\it b} and {\it y}  filters) and activity, that there is evidence for both the direct  \citep[e.g.][]{leanetal2005, krivovasolanki2008, krivova_rec2010,  balletal2011} and inverse  \citep[][]{harderetal2009, premingeretal2011} correlations. \cite{lockwoodetal2007} showed that while the solar 11-year cycle in chromospheric activity is vigorous relative to the solar analogs, its photometric variability is significantly smaller than indicated by the stellar data \citep[see also][and references therein]{shapiroetal2013_stars}. At the same time \cite{halletal2007a, halletal2009} showed that both chromospheric and photometric variabilities of 18 Scorpii, which is believed to be the closest bright solar analog \citep{analogs,petitetal2008},  are very similar to the solar values. Thus, even if the solar activity cycle is anomalous with respect to the majority of stars \cite[as was for example proposed by][]{stellar_dynamo} there are stars which exhibit solar-like chromospheric and photometric variabilities.

We note that  the comparison of stellar and solar variabilities is not straightforward.  The Sun-like stars have a broad range of chromospheric activities, bracketing the solar value. The more rapidly rotating ones may also have different latitudinal distributions of active regions on their surfaces \citep{polar1,granzer2002, berdyugina2005a} and, unlike the Sun, which is  observed from its near-equatorial plane, can  be observed from arbitrary directions \citep{schatten1993, knaacketal2001,luis2012}. All these factors affect the apparent variability of Sun-like stars and hinder the direct  comparison of solar and stellar photometric and chromospheric data series.

In this paper we present a model which allows us to understand how the solar photometric variability would manifest itself if the Sun had a different level of chromospheric activity. It also allows us to study the dependence of the photometric and chromospheric variability on the angle between the  stellar rotation axis and the line-of-sight, as well as on the latitudinal distribution of active regions on the surface.  It thus provides a tool for comparing the Sun to solar analogs, which may help to better understand the physical mechanisms of solar and stellar variability.

Our model is based on a simplified version of the SATIRE \citep[Spectral And Total Irradiance Reconstruction, see][]{fliggeetal2000, krivovaetal2003}  model of  solar irradiance variability expanded  from the Sun to stars with different activity levels (in particular to more active stars).
It allows us to understand  whether the observed patterns of stellar variability are compatible with the solar paradigm and can be explained by a simple extrapolation from the Sun.
Extrapolating from solar to stellar activity levels in order to explain the first observations of  \cite{lockwoodetal1992} was proposed by \cite{foukal1994}, although at a more quantitate level than we attempt in this paper.

In Sect.~\ref{sect:Sun} we give a brief overview of our understanding of solar irradiance variability.  In Sect.~\ref{sect:FF} we present empirical dependences of  the solar spot and facular disc area coverages on chromospheric activity and the datasets used to define them. {  In Sects.~\ref{sect:model} and Sect.~\ref{sect:BYS} we describe our model of stellar variabilty, based on the extrapolation of the empirical dependences introduced in Sect.~\ref{sect:FF}. In Sect.~\ref{sect:pat} we employ this model to explain the observed patterns of stellar variability.}
%{  We simulate the Brightness-Activity correlations for individual stars in Sect.~\ref{sect:BYS} and employ our model to explain the observed patterns of stellar variability in Sect.~\ref{sect:pat}}. 
Finally, we summarise our main results in Sect.~\ref{sect:conc}.

\section{Solar irradiance variability}\label{sect:Sun}
It has been known since antiquity, when first naked eye observations of sunspots were reported \citep[see discussion in][]{usoskin2008}, that the solar surface is inhomogeneous in brightness. The cause of this inhomogeneity is the magnetic field on the solar surface, which gives rise to sunspots and facuale, convection (granulation) and oscillations (p-modes). Whereas convection and oscillations are the dominant contributors to solar irradiance variability on time scales shorter than a day \citep{seleznyovetal2011}, the surface magnetic field is currently believed to be the main source of the solar irradiance  variability over time scales ranging from days to decades and centuries \citep{domingoetal2009,TOSCA2012, MPS_AA}. Modern  models \citep[e.g.][]{leanetal2005,krivovasolanki2008, shapiro_rec, fontenlaetal2011} attribute the variability of the solar irradiance to the imbalance between the contributions from dark (e.g. sunspots or pores) and bright (e.g. faculae or network) magnetic features in the solar atmosphere. The rotation of the Sun causes the irradiance modulation with a 27-day period, while changes in the overall magnetic activity (and consequently in the surface area coverage of active features)  lead to the 11-year activity cycle  and long-term changes in solar irradiance.

The variability of the solar irradiance affects the terrestrial climate \citep[e.g.][]{haigh2007,grayetal2010,TOSCA2012} and therefore has been under close scrutiny since the beginning of regular space-borne measurements. The TSI has been measured almost continuously since the launch of the NIMBUS 7 mission in 1978 \citep{hoytetal1992}. Its variability on the 27-day and 11-year time scales is relatively well constrained \citep{pmod_comp, balletal2012, MPS_AA}.

Measurements of the spectral solar irradiance (SSI) are not continuous and the SSI variability is significantly less constrained than the TSI variability. Most of the available datasets yield similar amplitudes and spectral profiles of the SSI variability on the solar rotational time-scale \citep{balletal2011,unruhetal2012,leandeland2012,TOSCA2012}. On longer time-scales the accumulated uncertainty in the instrumental degradation corrections becomes larger and renders the interpretation of the SSI measurements very difficult. Thus analysis of the 11-year activity cycle variability continues to produce surprises \citep{harderetal2009, woods2012, wehrlietal2013} and the magnitude of the 11-year SSI variability is still uncertain \citep{delandandcebula2012,TOSCA2012}.

%The picture of  solar variability on a centennial time scale is even more subtle as no direct measurements beside sunspot observations are available and the solar dynamo, which is believed to drive the variability, is still not fully understood \citep{dynamoLR}. There is an ongoing scientific controversy about the magnitude of the secular change in irradiance  \citep[see e.g.][]{Lockwood2011,schmidtetal2012,solankiandunruh2012}.

The substantial disagreements between existing SSI datasets lead to different atmospheric responses when they are used in climate models \citep{oberlanderetal2012, gerard_comp, shapiroetal2013,TOSCA2012} and hinder understanding of the Sun-Earth connection.

An alternative way to improve and deepen our understanding of solar variability on all time scales is to test and validate the available SSI models and measurements against observations of Sun-like stars. This approach is starting to attract attention and has already  been employed in the literature to constrain  solar variability \citep[][]{premingeretal2011, analysis, shapiroetal2013_stars}. The model presented below provides a quantitative basis for such a comparison.

\begin{figure}
\resizebox{\hsize}{!}{\includegraphics{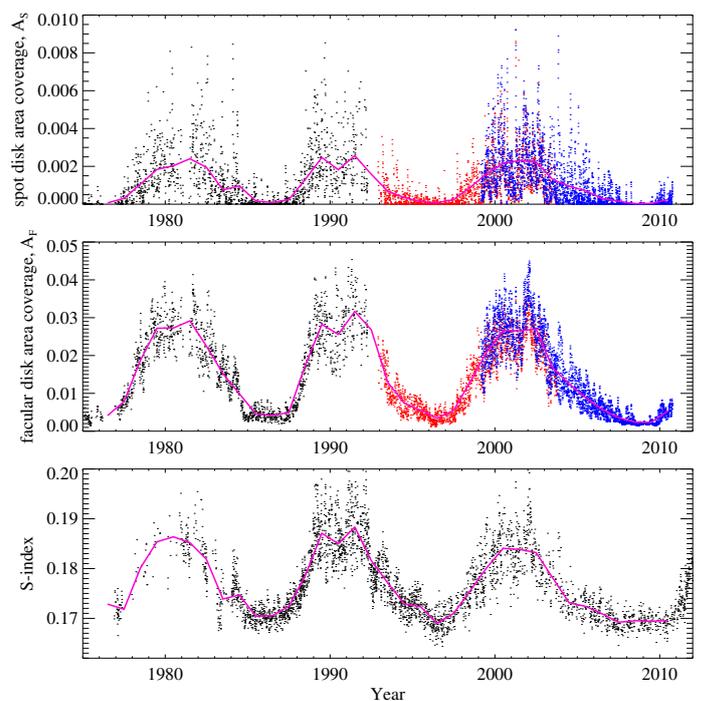}}
\caption{Disk area coverages by sunspots (upper panel) and solar faculae (middle panel)  as well as the S-index (lower panel). The daily disc area coverages are determined from the KP/512 (black dotes), KP/SPM (red dots), and SOHO/MDI (blue dots) full-disc continuum images and magnetograms. The daily S-index (black dots in the lower panel) is calculated from the Sac Peak Ca II data. Magenta curves are the annually averaged data.}
\label{fig:data}
\end{figure}

\section{Disk area coverage by active regions and chromospheric activity}\label{sect:FF}
During periods of high activity a noticeable part of the solar disc is covered by  faculae and spots. %(they can cover up to 5\% and 1\%, respectively).
At the same time, during periods of low activity the Sun can be completely spotless, e.g. there were 265 spotless days in 2008 according to Solar Influences Data Analysis centre\footnote{www.http://sidc.oma.be}. In this section we establish the connection between the fraction of the solar disc covered by active regions and the solar activity (characterised by the Ca II H and K emission, a widely used proxy in stellar activity research).

For this, we employ disc area coverages from the SATIRE-S model  \citep[with ``S'' standing for the satellite era, see][]{SATIRE}. SATIRE-S distinguishes three active components (together often referred to as active regions) representing spot umbra, spot penumbra, and faculae. The part of the solar disc not covered by these components is attributed to the quiet Sun. The disc area coverages are calculated from the full-disc continuum images and magnetograms obtained by the 512-channel diode array and spectromagnetograph at the National Solar Observatory Kitt Peak Vacuum Tower (KP/512, and KP/SPM, respectively) and by the Michelson Doppler Imager onboard the Solar and Heliospheric Observatory (SOHO/MDI). The detailed description of the datasets and the homogenizing procedure are presented in \cite{balletal2012}. To take into account that the small dark features, such as pores, are often not identifiable at the KP/512 continuum images we follow \cite{wenzleretl2006} and correct KP/512 spot coverages in that we multiply them by the factor 1.127.  Also, umbra and penumbra cannot be distinguished with the KP/512 data so that for all three instruments we obtain their sum (i.e. spot disc area coverage) instead and assume a fixed umbra to penumbra area ratio 1:4 \citep{wenzleretl2006,balletal2012} for the rest of the paper.

The spot and facular  disc area coverages for the Sun are plotted in Fig.~\ref{fig:data}. They both show a clear 11-year activity modulation. While the annual spot disc area coverage is close to zero during the solar minimum periods, the annual facular disc area coverage remains noticeably above the zero level even during the solar minima.

\begin{figure}
\resizebox{\hsize}{!}{\includegraphics{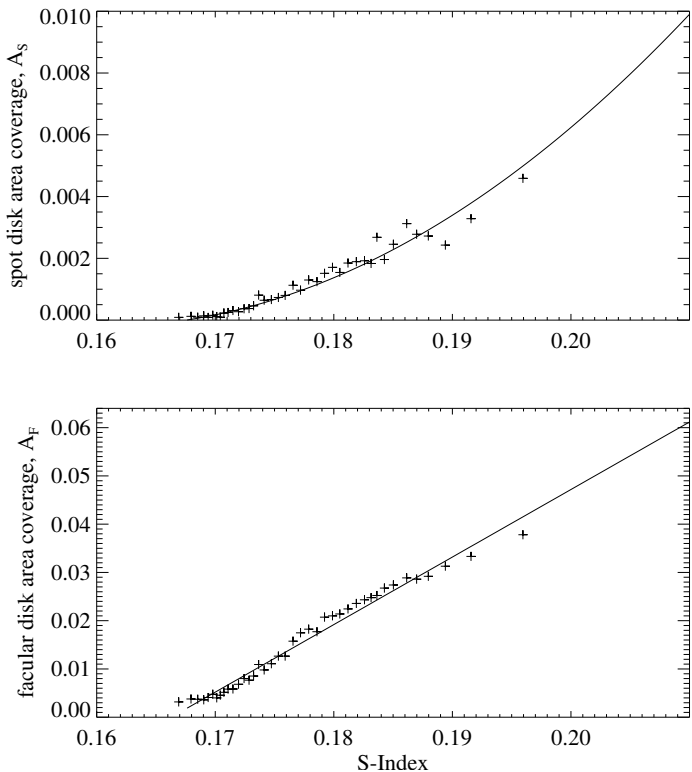}}
\caption{Dependence of the solar spot (upper panel) and facular (lower panel) disc area coverages on the S-index of chromospheric activity. The crosses correspond to the binned values, the thick curves are the least-square fit dependences (quadratic for spot disc area coverages, and linear for facular disc area coverages).}
\label{fig:dep}
\end{figure}

The disc integrated Ca II S-index of solar activity is proportional to the ratio between the summed flux in the Ca II H and K cores and the summed flux in two nearby continuum bands \citep[see][for a detailed discussion]{radicketal1998} and is often used as a proxy for solar and stellar chromospheric activity. However, the techniques employed for measurements of the solar and stellar Ca II indices are different. Additionally, there are multiple datasets of the solar Ca II index.
Therefore various conversion factors are usually employed  to connect different data. We use daily Sac Peak K-index K$_{\rm SP}$  \citep{CAII_SP} which can be transformed to monthly Kitt Peak K-index K$_{\rm SP}$ with a relationship $K_{\rm KP}=-0.01+1.1K_{\rm SP}$ \citep{whiteetal1998}.  The Kitt Peak K-index can, in turn, be transformed to the S-index: $S=1.53 K_{\rm KP} + 0.04$ \citep{whiteetal1992,radicketal1998}. The resulting solar S-index is plotted in the lower panel of Fig.~\ref{fig:data}.

To establish the dependence of the disc area coverage by active regions on the S-index we consider all days for which  simultaneous measurements of the S-index and  disc area coverages by spots and faculae are available. We sort these days according to the S-index and split the resulting monotonous series of the S-index into bins containing 58 days 
%(the effect of the bin size on the dependence of the disc area coverages  on the S-index is considered in Appendix~\ref{app:dev}). 
Then we calculate the mean value of the S-index and disc area coverages for every bin. 

Figure~\ref{fig:dep} illustrates the relationships between the binned disc area coverage and the S-index.  One can see that while facular disc area coverage increases linearly  with the S-index, the spots display rather a quadratic relationship \citep[see also][]{foukal1998, solankiandunruh2012}. Thus the ratio between spot and facular disc area coverages increases with activity.  By applying a least-squares fit and prescribing the value of the error of the mean disc area coverage to the  standard deviation, we found the following dependence for the sunspot disc area coverage $A_S$:
\begin{equation}
A_S (S)=(0.105  \pm 0.011)-(1.315 \pm 0.130)S + (4.102 \pm 0.370)S^2,
% A_S (S)=(0.097 \pm 0.011)-(1.218 \pm 0.124)S + (3.812 \pm 0.352)S^2,
\label{f_spot}
\end{equation}
and for facular disc area coverage $A_F$:
\begin{equation}
A_F (S)=-(0.233 \pm 0.002)+(1.400 \pm 0.010)S. 
\label{f_fac}
\end{equation}
Here the errors correspond to 1$\sigma$-uncertainty.  Note that all terms are significant at the 9$\sigma$ level.  On the contrary, if instead of the linear dependence in Eq.~(\ref{f_fac}) we use a quadratic relationship,  the quadratic term is  insignificant at the 2$\sigma$ level.

We note that the S-index and the disc area coverages are strongly variable on the 27-day solar rotation time scale. If instead of the binned values we used time averages (e.g. annual values), all information about the variability on the solar rotation time scale would be lost and additionally the uncertainty of the mean disc area coverages would be larger. This would hinder our analysis.

The minimum  annual value of the S-index based on the Sac Peak Ca II data over the last three solar activity cycles was reached  in 1996 and equals 0.169. According to Eqs.~(\ref{f_spot})--(\ref{f_fac}) this results in $A_S \approx 0.003\%$ and $A_F \approx 0.36\%$ at that time. The maximum annual value of 0.188 was reached in 1991, which corresponds to  $A_S \approx 0.28\%$ and $A_F \approx 3\%$.

\section{Model: calculations of the photometric brightness and chromospheric activity}\label{sect:model}
In this section we describe the model which allows us to establish the link between the stellar chromospheric activity (as traced by the S-index) and the photometric brightness and to explain the observed patterns of stellar variability. 

Our model is conceptually an extrapolation of a simplified version of the SATIRE model for solar irradiance variability  to stars with different levels of  chromospheric activity and, consequently, different coverages by active regions.  Following the SATIRE approach, we decompose the stellar atmosphere into the four components: quiet regions, faculae, spot umbra, and spot penumbra. We also employ the SATIRE spectra of these components \citep[see][for the detailed description]{sat_spectra}, which are known to  conform with the disc area coverages described in Sect.~\ref{sect:FF} \citep[see e.g.][]{balletal2011}. This ensures the proper representation of the solar variability by our model \citep[see also][]{solankiandunruh2012}.  We note that when applied to the Sun our model leads to slightly different  results than the model  presented in \cite{knaacketal2001}, who studied the dependence of the spectral solar irradiance and the S-index on the angle between the direction to the observer and stellar rotational axis (hereafter stellar inclination). The reason for this is that \cite{knaacketal2001} used a slightly different spot model atmosphere and employed a simplified approach to calculate the dependence of the  S-index on inclination.

The main goal of our approach is to extrapolate the dependences established in Sect.~\ref{sect:FF} (Eqs.~\ref{f_spot}--\ref{f_fac}) to higher activity levels and to use them to calculate stellar spot and facular disc area coverages as functions of the S-index. This allows simulations of a magnetically active Sun by filling its surface with an increasing fraction of sunspots and faculae. Assuming a fixed umbra to penumbra area ratio (see Sect.~\ref{sect:FF}), the spot disc area coverage can be decomposed into umbral and penumbral coverages. The impact of the uncertainties in the extrapolation of disc area coverages on our results is discussed in Appendix~\ref{app:dev}.

In this paper we focus on the variability on time scales of stellar activity cycles and only use the seasonally averaged stellar data.
Thus we can assume that active regions are uniformly distributed over the activity belts  whose positions on the stellar surface do not change with time, which is an additional simplification that we make due to the unknown shape of the butterfly diagram in stars. 
This allows us to calculate the change of stellar photometric brightness due to active regions, using  the full disc area coverages given by Eqs.~(\ref{f_spot})--(\ref{f_fac}) and assuming that the contrasts between the active and quiet regions are the same as for the Sun. Our approach thus treats stars as hypothetical Suns whose coverage by active regions may or may not differ from that of the Sun. The use of solar relationships (Eqs.~\ref{f_spot}--\ref{f_fac}) means that we test how well the solar paradigm reproduces the behavior of more active stars. The use of solar model atmospheres implies that an accurate comparison can be carried out only with main sequence stars of a similar spectral type.

\begin{figure*}
\resizebox{\hsize}{!}{\includegraphics{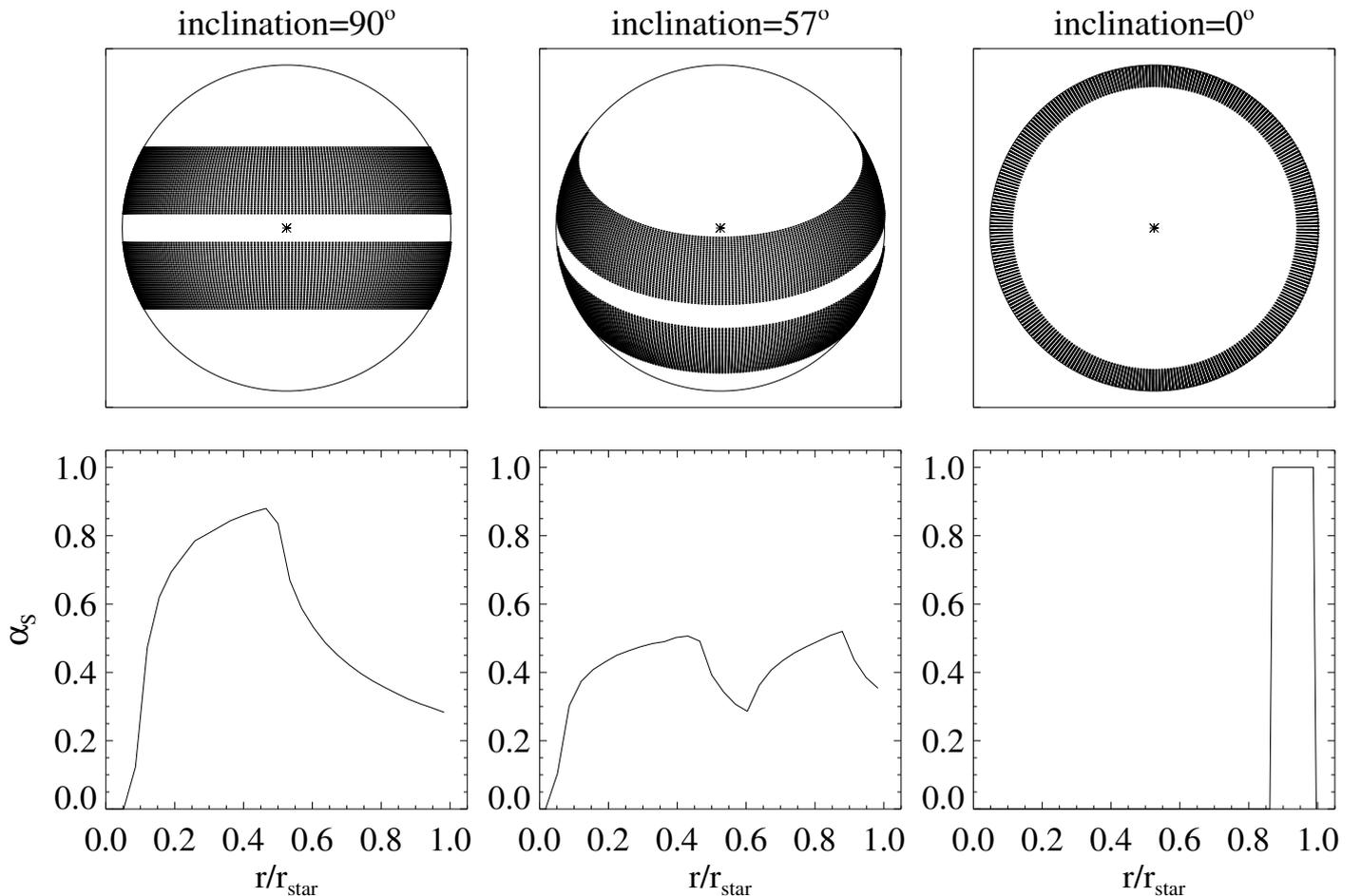}}
\caption{Upper panels: Visible projections of two bands between latitudes $\pm 5^\circ$  and $\pm 30^\circ$ on the stellar disc for three values of the inclination. The bands correspond to the solar distribution of spots, employed in this paper. The star sign indicates the position of the disc centre. Lower panels:  the fractional coverage of the stellar disc by spots plotted as functions of the impact parameter $r/r_{\rm star}$ (which is 0 at the disc centre and 1 at the limb). The fractional coverage is calculated assuming that both bands plotted in the upper panels are completely covered by spots. }
\label{fig:ff_solar}
\end{figure*}

In general, the latitudinal  distribution of active regions on the stellar disc might be different from that of the Sun. The situation is further complicated by the fact that while the Sun  is basically observed from its equatorial plane (the angle between the solar equator and ecliptic is $\sim 7.25^\circ$ and will be neglected in all further calculations),  stellar inclinations may vary from 0$^\circ$ (the star is observed along the rotational axis) to 90$^\circ$ (the star is observed from its equatorial plane, which is similar to observing the Sun from the ecliptic) and is usually poorly known. This affects both the photometric brightness and the measured chromospheric activity.

Thus the connection between the photometric brightness and the observed S-index is additionally affected by the latitudinal  distribution of active regions and the stellar inclination. In Sects.~\ref{sect:model.BY}  and \ref{sect:model.S} we show how to calculate photometric and chromospheric variability assuming a solar distribution of active regions, i.e. spot bands between 5$^\circ$ and 30$^\circ$ and facular bands between 5$^\circ$ and 40$^\circ$ \citep{knaacketal2001}. The opposite case of a polar distribution of active regions will be considered in Sect.~\ref{sect:model.polar}.

\subsection{Photometric brightness}\label{sect:model.BY}
The spectral flux can be decomposed into the quiet and active components:
\begin{equation}
F(\lambda)=F_{\rm quiet} (\lambda)+ F_{\rm active}(\lambda).
\label{eq:sum}
\end{equation}
The quiet component $F_{\rm quiet} (\lambda)$ represents the hypothetical case of the stellar disc without any magnetic features on the surface:
\begin{equation}
F_{\rm quiet} (\lambda) = \int_{\rm disc} I_Q(\lambda, {\vec r})  \, d \Omega.
\label{eq:quiet_prepare}
\end{equation}
Here the integration is done over the visible stellar disc, $I_Q (\lambda, {\vec{r}})$ is the intensity from the quiet stellar component along the direction $\vec r$, and $d \Omega$ is a differential of the solid angle around the direction $\vec r$. SATIRE employs 1D models of the solar atmosphere so that the emergent intensity $I_Q (\lambda, {\vec{r}})$ from a given component of the atmosphere depends only on the  wavelength and the angle between the direction to the disc centre of the star and $\vec r$.

The emergent intensity is often written as a function of the cosine of the angle between the  direction to the observer and the local stellar radius, $\mu$. Then Eq.~(\ref{eq:quiet_prepare}) can be rewritten as
\begin{equation}
F_{\rm quiet} (\lambda) =\int_0^1  I_Q(\lambda, \mu) \, \omega (\mu) \, d \mu ,
\label{eq:quiet}
\end{equation}
where the weighting function $\omega (\mu) =2 \pi \mu (r_{\rm star}/d_{\rm star})^2$  (here $r_{\rm star}$ is the stellar radius and $d_{\rm star}$ is the distance between the star and the observer) represents the transformation between $d \Omega$ and $d \mu$.

We note that starting from 1974 (when KP/512 became available, see Sect.~\ref{sect:FF}) the solar irradiance was never equal to $F_{\rm quiet} (\lambda)$ and was affected by the active component, whose contribution is given by
\begin{equation}
F_{\rm active}(\lambda) =  \int_0^1   \sum_k   \alpha_k (\mu) \, \left ( I_k(\lambda, \mu) - I_Q(\lambda, \mu) \right ) \omega (\mu) \, d \mu.
\label{eq:active}
\end{equation}
Here  $ I_k(\lambda, \mu) - I_Q(\lambda, \mu)  $ is the wavelength- and angle-dependent contrast between the active component $k$ and quiet stellar regions. In the ultraviolet and visible spectral ranges it is positive for  faculae  (i.e. faculae cause a brightness excess)  and negative for umbra and penumbra (i.e. spots cause a brightness deficit). The functions $\alpha_k (\mu)$ give the fractional coverage of the {\it ring} around the stellar disc centre with a radius corresponding to $\mu$ by  the component $k$ (e.g. $\alpha_k (1)$  and   $\alpha_k (0)$ represent the stellar disc centre and limb, respectively). Integration over $\mu$ gives $A_k$,   the fractional coverage of the whole visible stellar disc by the component $k$:
\begin{equation}
A_k={\int_0^1 \alpha_k(\mu) \omega (\mu) \, d \mu} / {\Omega_{\odot}} ,
\label{eq:geom}
\end{equation}
where  $ \Omega_{\odot}={\int_0^1 \omega(\mu) d \mu}$ is the solid angle subtended by the star.

%We note that the disc area coverages $A_k$ correspond to the fractional coverage of the whole visible stellar disc (which is the projection of the visible stellar spherical surface into the line-of-sight). They are identical to the disc area coverages  derived from Eqs.~\ref{f_spot}--\ref{f_fac}.
%The connection between $\alpha_k$ and $A_k$ depends on the location and shape of the activity belts as well as on the inclination.
 
While the stellar inclination does not directly enter Eqs.~(\ref{eq:active})-(\ref{eq:geom}), it affects the area coverages  $\alpha_k (\mu)$, which depend on the inclination and the latitudinal distribution of active regions. This is illustrated in Fig.~\ref{fig:ff_solar}, which represents area coverages of spots for three inclinations. Note that for convenience in the lower panels, the area coverages are plotted as functions of the impact parameter $r/r_{\rm star} = \sqrt{1-\mu^2}$. For all three cases the disc centre is located outside of the visible activity belts, so that $\alpha_k=0$ for  $r/r_{\rm star}=0$ (or $\mu=1$). The value of $\alpha_S$ becomes positive when the corresponding ring starts to intersect the visible activity belts. The sudden decrease of $\alpha_S$ at $r/r_{\rm star}=0.996$ in the lower right panel of Fig.~\ref{fig:ff_solar} is caused by the presence of the spot-free band between $0^\circ$ and $5^\circ$. 
%which is free of spots. For a star which is observed from its rotation axis (zero inclination), this band projects into the near-limb region (which is not visible in the scale of the upper panel in Fig.~\ref{fig:ff_solar}) so that $\alpha_k=0$ for  $r/r_{\rm star}=1$. 
A similar figure for the polar distribution of active regions (see Sect.~\ref{sect:model.polar}) is given in the Online Material (Fig.~\ref{fig:ff_polar}). One can see that the change of the inclination or the  latitudinal  distribution of active regions might completely reshape the $\alpha_k(r/r_{\rm star})$ dependences and consequently the $\alpha_k (\mu)$ dependences used in  Eqs.~(\ref{eq:active})-(\ref{eq:geom}).

\begin{figure}
\resizebox{\hsize}{!}{\includegraphics{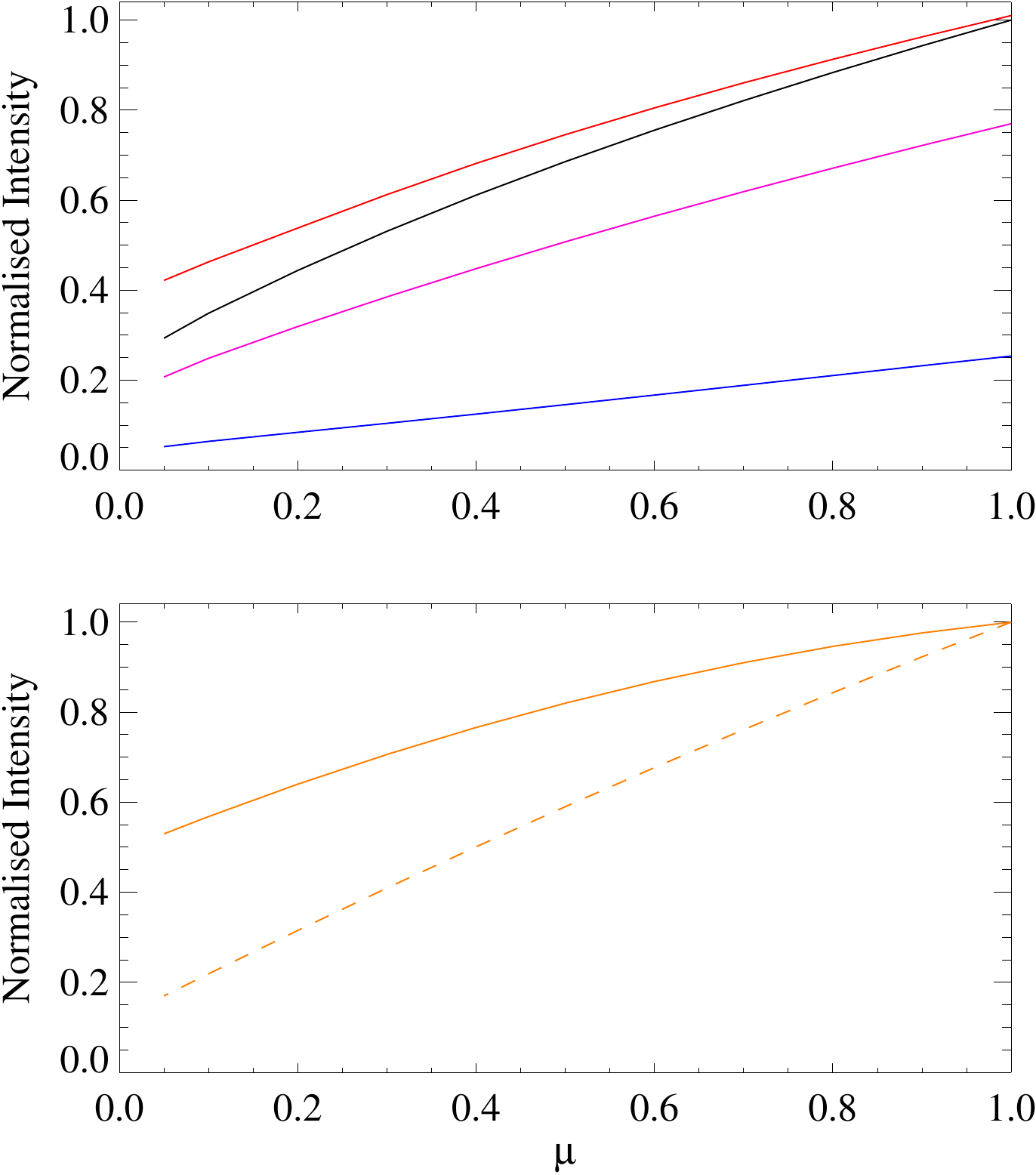}}
\caption{Upper panel: centre-to-limb variation of the Str{\"o}mgren $(b+y)$/2 intensity calculated for the quiet Sun, faculae, penumbra, and umbra (black, red, magenta, and blue curves) models. Plotted are the ratios to the quiet Sun's intensity at the disc centre. {  Lower panel: quadratic polynomial  parameterisation of the intensity centre-to-limb variation at the Ca II K line and the nearby continuum  (solid and dashed orange curves, respectively)  according to \cite{CAII_CLV}.}  Both intensities are normalised to unity at the disc centre.}
\label{fig:CLV}
\end{figure}

The contribution of a given ring with fixed $\mu$ value to $F_{\rm active}$ depends on the fractional coverages $\alpha_k (\mu)$, weighting coefficient $\omega (\mu)$ and  centre-to-limb variations (CLV) of the intensity emergent from different atmospheric components. The {  calculated} CLV of the spectral intensity averaged over the Str{\"o}mgren $(b+y)$/2 profile are shown in the upper panel of Fig.~\ref{fig:CLV} for the four atmospheric components employed in SATIRE. The contrast between the quiet regions and faculae increases steeply from the disc centre towards the limb. This implies that enhancement of the irradiance caused by the faculae also increases from the disc centre towards the limb. At the same time the relative umbral and penumbral contrasts are only barely affected by the CLV.

In the case of the solar distribution of active regions Eqs.~(\ref{f_spot})--(\ref{f_fac}) define the disc area coverages $A_k$ (however, some caution is needed as the S-index which enters these equations corresponds to observations from the equatorial plane and is, thus, different from the S-index measured on a star with another inclination of its rotation axis, see Sect.~\ref{sect:model.S}).  Eq.~(\ref{eq:geom})  makes it possible to convert the $A_k$ to a fractional coverages $\alpha_k (\mu)$, which enter Eq.~(\ref{eq:active}). Hence, the set of Eqs.~(\ref{f_spot})--(\ref{eq:geom}) allows calculations of the photometric brightness as a function of the S-index for an arbitrary stellar inclination.  

\subsection{Chromospheric activity}\label{sect:model.S}
The S-index is proportional to the ratio of the flux in the cores of the Ca II H and K lines to the flux in the nearby continuum. \cite{CAII_CLV} showed that the centre-to-limb variation of the Ca II K emission  \citep[which hereafter will be used as a representation of the summed emission in the H and K lines, see][]{radicketal1998} is the same for the quiet Sun and faculae. This was confirmed by later measurements \citep[see e.g.][]{ermollietal2007, ermollietal2010}.  Thus one can write
\begin{eqnarray}
&& I_{\rm HK}^F(\mu)=I_{\rm HK}^F(1) \cdot C_{\rm HK}(\mu),  \nonumber \\
&& I_{\rm HK}^Q(\mu)=I_{\rm HK}^Q(1) \cdot C_{\rm HK}(\mu), \\
&& I_{\rm cont}^Q(\mu)=I_{\rm cont}^Q(1) \cdot C^Q_{\rm cont}(\mu). \nonumber
\end{eqnarray}
Here $I_{\rm HK}^Q$ and $ I_{\rm HK}^F$  are the Ca II H and K intensities (hereafter, HK intensity) from the quiet Sun and from faculae, respectively, and $I_{\rm cont}^Q$ is the quiet Sun's intensity in the nearby continuum. Functions $ C_{\rm HK}(\mu)$ and $C_{\rm cont}^Q(\mu)$ describe the CLV of the HK and continuum fluxes.  The HK flux and its CLV are strongly affected by the effects of non-local thermodynamical equilibrium \citep[see e.g.][]{ermollietal2010}, which are not taken into account in the spectra employed by the SATIRE model \citep{sat_spectra}. Therefore, instead of taking the theoretical CLV resulting from SATIRE we employ {  quadratic polynomial  parameterisations (as a function of the heliocentric angle cosine $\mu$) of the CLV from \cite{CAII_CLV}, who employed several observational datasets to define them. The parameterisations for the Ca II K line core and the nearby continuum are shown in the lower panel of Fig.~\ref{fig:CLV}}.

For a given inclination $i$, the S-index can be written as 
\begin{eqnarray}
S_i = &&\frac{I_{\rm HK}^Q(1) \cdot \int_0^1 \alpha_Q(\mu) \cdot C_{\rm HK}(\mu) \, \omega (\mu)  \, d \mu} {I_{\rm cont}^Q(1) \cdot \int_0^1 \left (\alpha_Q(\mu)+\alpha_F(\mu) \right ) C^Q_{\rm cont}(\mu) \, \omega (\mu)  \, d \mu}      \nonumber \\  
&& \,\,\,\,\,\, \,\,\,\,\,\,+ \frac{I_{\rm HK}^F(1) \cdot \int_0^1 \alpha_F(\mu) \cdot C_{\rm HK}(\mu) \, \omega (\mu)  \, d \mu  }{I_{\rm cont}^Q(1) \cdot \int_0^1 \left (\alpha_Q(\mu)+\alpha_F(\mu) \right ) C^Q_{\rm cont}(\mu) \, \omega (\mu)  \, d \mu.}
\label{eq:S}
\end{eqnarray}

%\begin{eqnarray}
%S_i = &&\frac{I_{\rm HK}^Q(1) \cdot \int_0^1 \alpha_Q(\mu) \cdot C_{\rm HK}(\mu)  \,  \omega(\mu) \, d \mu} {I_{\rm cont}^Q(1) \cdot \int_0^1  C_{\rm cont}^Q(\mu) \, \omega(\mu) \, d \mu}      \nonumber \\  
%&& \,\,\,\,\,\, \,\,\,\,\,\,+ \frac{I_{\rm HK}^F(1) \cdot \int_{0}^1 \alpha_F(\mu) \cdot C_{\rm HK}(\mu)   \,  \omega(\mu) \, d \mu  }{I_{\rm cont}^Q(1) \cdot \int_{0}^1  C_{\rm cont}^Q(\mu) \, \omega(\mu) \, d \mu.}
%\label{eq:S}
%\end{eqnarray}
The first and second terms in Eq.~(\ref{eq:S}) give the contribution of the quiet stellar regions and faculae, respectively.  In the continuum the contrast between the quiet regions and faculae is small \citep[unlike the contrast in the H and K lines, see Fig.~1 from ][]{CAII_CLV} so that it is neglected in the denominator of Eq.~(\ref{eq:S}). Additionally we assume for simplicity that both umbral and penumbral regions are completely dark and simply dilute the continuum and HK fluxes.

We note that the dependence of the S-index on inclination and latitudinal  distribution of active regions enters Eq.~(\ref{eq:S}) via the area coverages $\alpha_Q(\mu)$ and $\alpha_F(\mu)$, exactly in the same way as it enters Eq.~(\ref{eq:active}) for $F_{\rm active}$ (see discussion in Sect.~\ref{sect:model.BY} as well as Figs.~\ref{fig:ff_solar} and \ref{fig:ff_polar}). Therefore the measured S-index  depends on the inclination and the same star may have different S-indices if observed from different directions. 

Using the fact that $\alpha_S+\alpha_F+\alpha_Q=1$,  Eq.~(\ref{eq:S}) may be rewritten as follows:
\begin{equation}
S_i = S^Q \cdot \frac{I_2-I_3-I_4}{I_1-I_5}+S^F \cdot \frac{I_3}{I_1-I_5},
\label{eq:S_i}
\end{equation}
where 
\begin{eqnarray}
I_1 \equiv \int_0^1 C_{\rm cont}^Q\!\!\!\!\!\!&&(\mu) \, \omega (\mu) \, d \mu; \,\,\,\,\,\, \,\,\,\,\,\,
I_2 \equiv \int_0^1  C_{\rm HK}(\mu) \, \omega (\mu) \,  d \mu \nonumber ; \\
&& I_3 \equiv \int_0^1 \alpha_F(\mu) \, C_{\rm HK}(\mu) \,  \omega (\mu) \, d \mu \nonumber ; \\
&& I_4 \equiv \int_0^1 \alpha_S(\mu) \cdot C_{\rm HK}(\mu) \, \omega (\mu)  \, d \mu \nonumber ;  \\
&& I_5 \equiv \int_0^1 \alpha_S(\mu) \cdot C^Q_{\rm cont}(\mu) \, \omega (\mu)  \, d \mu.  
\label{eq:int}
\end{eqnarray}
and
\begin{equation}
S^Q \equiv I_{\rm HK}^Q(1)/I_{\rm cont}^Q(1) ; \,\,\,\,\,\, \,\,\,\,\,\,
S^F \equiv I_{\rm HK}^F(1)/I_{\rm cont}^Q(1).
\label{S_def}
\end{equation}

The integrals $I_1$ and $I_2$ in Eq.~(\ref{eq:S_i}) depend neither on inclination, nor on activity.  Since $A_F$ is a  linear function of S$_{90}$ (which is the S-index observed from the equatorial plane, see Eq.~\ref{f_fac}) and we assume a fixed latitudinal distribution of active regions,  the area coverage $\alpha_F(\mu)$ increases linearly with S$_{90}$. Hence the integral $I_3$  is  a linear function of S$_{90}$.

Let us consider a case of a star with low level of mean chromospheric activity observed from its equatorial plane (e.g. the Sun) with measured S-index equal to  S$_{90}^{\rm obs}$. For low levels of magnetic activity the integrals $I_4$ and $I_5$ are negligibly small so that the sunspots do not have any effect on the S-index \citep[see][]{knaacketal2001} and $S_i$ is a linear  function of S$_{90}^{\rm obs}$.  The S$_{90}^{\rm obs}$ value can be employed to calculate the disc area coverages $A_k$ (using Eqs.~\ref{f_spot}--\ref{f_fac}) which then can be converted into the area coverages $\alpha_k(\mu)$.   Knowing the area coverages one can calculate the integral $I_3$ (see Eq.~\ref{eq:int}) and employ Eq.~(\ref{eq:S_i}) to calculate S$_{90}^{\rm calc}$. If our approach is self-consistent then the calculated S$_{90}^{\rm calc}$ value must be equal to the S$_{90}^{\rm obs}$ value employed to drive our calculations. In other words,  Eqs.~(\ref{eq:S_i})-(\ref{S_def}) define the dependence of the calculated S$_i$-index on disc area coverages, while Eq.~(\ref{f_fac}) defines the inverse dependence of disc area coverages on the S$_{90}$-index. For a star observed from its equatorial plane these dependences must agree with each other. As discussed above, the right side of Eq.~(\ref{eq:S_i}) is a linear function of S$_{90}^{\rm obs}$, so that this equation expresses  S$_{90}^{\rm calc}$ in terms of  S$_{90}^{\rm obs}$. Therefore, the condition 
$S_{90}^{\rm calc}= S_{90}^{\rm obs} $ is fulfilled if the parameters $S^Q$ and $S^F$ take on a particular set of values.

For the solar distribution of active regions this condition  is fulfilled if S$^Q$= 0.137 and S$^F$=0.711. These parameters are the values of the S-index from the quiet regions and the faculae observed at the disc centre.  The ratio between $S^Q$ and $S^F$ values agrees well with the Ca II K  line profiles plotted in Fig.~1 of \cite{CAII_CLV}.  This means that the disc area coverages given by Eqs.~(\ref{f_spot})--(\ref{f_fac})   conform with the CLV of HK flux employed in our model.
The  linear dependence between facular disc area coverage and activity (see Eq.~\ref{f_fac})  implies that the $S^Q$ and $S^F$ parameters do not depend on the activity, i.e. the increase of the activity level affects the disc area coverages but not the structures of the individual components.

We note that the solar S-index of chromospheric activity which enters Eqs.~(\ref{f_spot})--(\ref{f_fac}) is measured from the equatorial plane (hereafter $S_{90}$). Consequently, these equations cannot be directly applied to calculate the disc area coverages by active regions from the $S_i$-index  measured from an arbitrary inclination $i$ and the $S_i$-index must first be converted into the the $S_{90}$-index. Eq.~(\ref{eq:S_i}) provides a simple way to make such a conversion.

\subsection{Different distributions of active regions}\label{sect:model.polar}
Already the first results of the Doppler imaging technique \citep[see e.g.][and references therein]{berdyugina2005a} pointed to the existence of stellar spots at high latitudes, reaching even to the poles for stars with deep convection zones (in complete contrast to the solar case, where the spots are rarely observed above 30$^\circ$ latitude). To explain the existence of high-latitude spots, \cite{polar1} suggested that while the rise of magnetic flux tubes through the convection zone in slowly rotating stars is controlled by the buoyancy force, the Coriolis force starts to be important for rapidly rotating stars. An alternative explanation, based on meridional flow, was later proposed by \cite{SchrijverandTitle2001}.  As a result, active features are distributed in the equatorial belts for the slowly rotating stars (like the Sun) and at higher latitudes for the rapidly rotating stars. The distribution of  active features with latitude for the Sun-like stars considered in this paper is generally unknown. 

As discussed above, the equations given in Sect.~\ref{sect:model.BY}~and~\ref{sect:model.S}  are valid for an arbitrary latitudinal distribution of  active regions.
All geometrical information (i.e. inclination of the star and distribution of active regions) enters these equations via the area coverages $\alpha_i (\mu)$. The modification of the latitudinal  distribution of active regions  affects the area coverages and consequently the contribution of the active regions to the photometric brightness (via the $F_{\rm active}$ term, see Eqs.~\ref{eq:sum}~and~\ref{eq:active}) and the S-index (via the $I_3$ integral, see Eqs.~\ref{eq:S_i}~and~\ref{eq:int}).

The main sophistication is that the S$_{90}$-index, which is used to parameterise  the spot and facular disc area coverages in Sect.~\ref{sect:FF}, corresponds to the solar distribution of active regions. The redistribution of the active regions affects S$_{90}$ even if the disc area coverages remain unchanged, so Eqs.~(\ref{f_spot})--(\ref{f_fac})  are only applicable for the solar distribution of active regions. 

To take this into account we first consider a star with  a solar distribution of active regions. For such a star we can employ Eqs.~(\ref{f_spot})--(\ref{f_fac})~and~(\ref{eq:geom}) to calculate the area coverages $\alpha_k (\mu)$ as  functions of  the parameter S$_{90}^{\rm S}$ (hereafter,  index ``S'' represents the solar distribution of active regions) and equations from Sect.~\ref{sect:model.BY}~and~\ref{sect:model.S}  to calculate S$_i^{\rm S}$ and $F_i^{\rm S}$. Then we rearrange the location of active regions according to the different  distribution and recalculate area coverages  $\alpha_k (\mu)$,  keeping the part of the stellar {\it surface} covered by each component {\it k}  (i.e.  $\int_{0}^1 \alpha_k (\mu) \, \omega (\mu) /   \mu   \, d \mu  $) unaffected. These new area coverages allow us to recalculate  the  chromospheric activity $S_i$    and photometric brightness $F_i$ for an arbitrary distribution of active regions.

%As discussed above, the right-hand side of Eq.~(\ref{eq:S_i}) is a linear function of S$_{90}^{\rm S}$. This implies that the connection between $S_i^{\rm P}$ and S$_i^{\rm S}$ is also linear. %, for example if $i=90^{\circ}$ then $S_{90}^{\rm P} = 0.53 S_{90}^{\rm S} + 0.078$.

In addition to the solar case we also consider the homogeneous distribution of active regions on a stellar surface as well as the polar distribution when spots and faculae  are gathered  in two polar caps (North and South) with latitude $>$ 45$^\circ$. The range of latitudes for the polar distribution is larger than for solar  because of the smaller area covered by a ring of given latitude width at high latitudes compared to low latitudes. Even with the polar caps reaching so far down they offer less surface area, covering only 34\% of the stellar surface compared to the 56\% covered by the solar activity belts. 

We expect that solar, polar, and homogeneous distributions of active regions  should encompass the behavior for most kinds of latitudinal distributions.

\subsection{The S-index as a proxy of stellar magnetic activity}\label{S_proxy} 
\begin{figure}
\resizebox{\hsize}{!}{\includegraphics{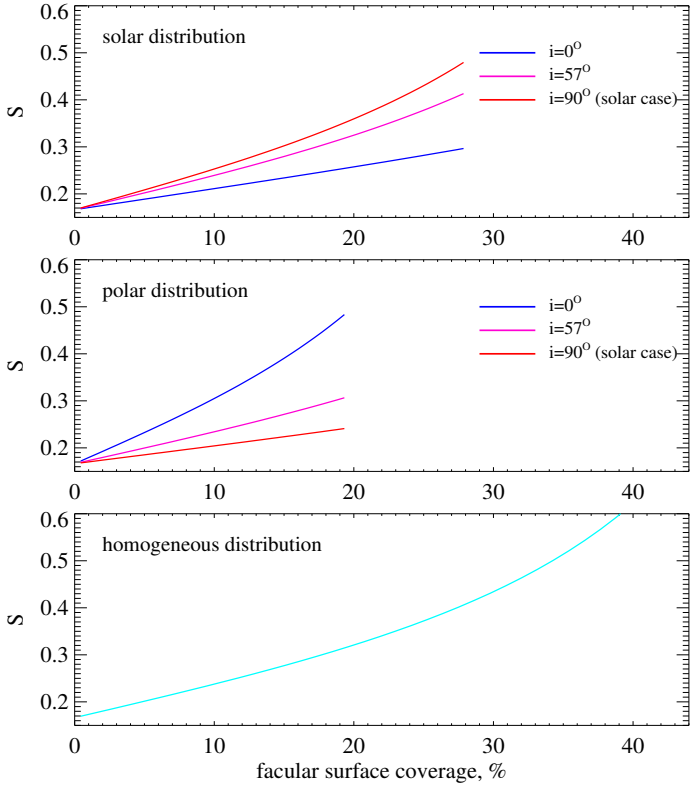}}
\caption{{  Simulated values of the S-index as they would be measured for inclinations of 90$^{\circ}$ (red curves), 57$^{\circ}$ (magenta curves), 0$^{\circ}$ (blue curves)  vs. facular surface coverage for solar,  polar, and homogeneous distributions of active regions.} The maximum possible facular surface coverage (i.e. when faculae and spots fully cover the activity belts/caps or entire stellar surface) is about 28\%, 19\%, and 47\% for  the adopted solar, polar, and homogeneous distributions respectively.}
\label{fig:FF_S}
\end{figure}

In Fig.~\ref{fig:FF_S} we plot the S-index as a function of facular {\it surface} (i.e. hemispheric) coverage for solar, polar, and homogeneous distributions of active regions and for three inclinations: $i=0$, $i=57^{\circ}$ \citep[the mean inclination which is often used in the literature to characterize the intermediate case, see e.g. ][]{schatten1993, knaacketal2001}, and $i=90^{\circ}$. In the case of a homogeneous distribution of active regions, the S-index does not depend on the inclination so that  the dependence is also the same for  different inclinations.

Integrals $I_4$ and $I_5$ are proportional to the spot surface coverage, which is a quadratic function of the facular surface coverage. Therefore the S-index is not a linear function of the facular surface coverage. 
%To illustrate the effect of spots on the S-index in Fig.~\ref{fig:FF_S} we also plot the linear dependences calculated neglecting $I_4$ and $I_5$ integrals in the Eq.~\ref{eq:S_i}.  
The effect of spots on the S-index is equivalent to the reduction of the  contribution of quiet regions to the continuum and HK emissions. This contribution is more important for the continuum emission as the HK emission is dominantly affected by the faculae. Thus, the spots diminish the continuum flux stronger than the HK flux and lead to an increase of the S-index. 

We note that the facular {\it surface} coverage is different from the facular {\it disc}  coverage $A_F$. The former is calculated for the visible part of the stellar surface (which is a hemisphere) and proportional to $\int_{0}^1 \alpha_k (\mu) \,   \, d \mu  $, while the latter is calculated for the visible stellar disc and proportional to $\int_{0}^1 \alpha_k (\mu) \, \mu  \, d \mu  $ (see Eq.~\ref{eq:geom}). All distributions of active regions considered in this paper are centrally symmetric with respect to the centre of the star (which we deem to be a sphere). Thus the surface coverages of the visible front-side of the star are equal to the surface coverages of the invisible far-side of the star. 

Stars with the same level of facular surface coverage are expected to have the same amount of magnetic flux emerging on the surface, but can have different facular  disc  coverages $A_F$ and measured S-indices (depending on the inclination and latitudinal  distribution of active regions).
Consequently, if two stars have the same S-indices it does not necessarily mean that they have the same amount of magnetic flux.  Thus, the S-index cannot be considered as an unequivocal proxy for stellar magnetic activity.

\begin{figure}
\resizebox{\hsize}{!}{\includegraphics{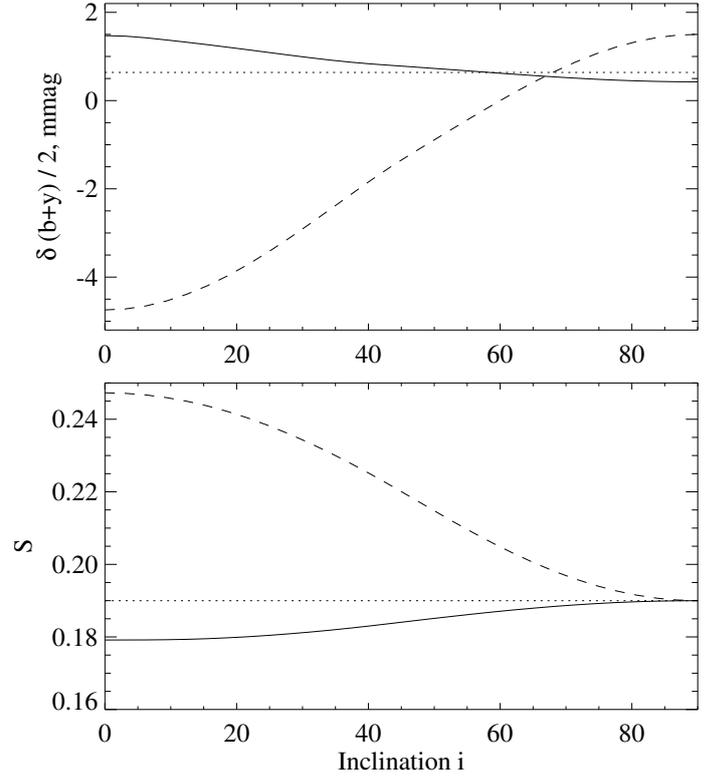}}
\caption{{  Simulated values of the} photometric brightness change $\delta (b+y)/2$    (upper panel) and S-index (lower panel) vs. stellar inclination $i$. Solid curves correspond to the solar distribution of active regions, dashed to polar distribution,  and dotted  to homogeneous distribution. {  The coverage by faculae and spots for all three distributions is chosen so that the S-index observed from the stellar equatorial plane ($i=90^{\circ}$) equals 0.19.}}
\label{fig:S_BY}
\end{figure}

\begin{figure}
\resizebox{\hsize}{!}{\includegraphics{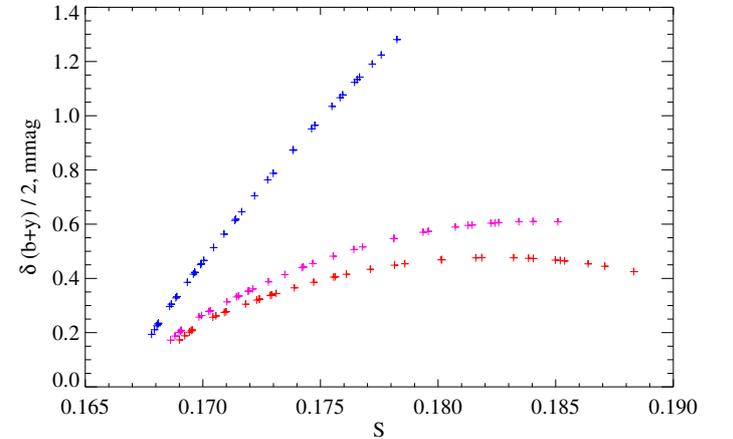}}
\caption{{  Simulated values of the} photometric brightness change $\delta (b+y) / 2$  vs. S-index for the Sun observed from three different directions. Plotted are the annual values {  as they would be measured} from the equatorial plane (red crosses), a 57$^{\circ}$ inclination (magenta crosses), and from the solar rotational axis (blue crosses). }
\label{fig:Sun_angles}
\end{figure}

\section{Model: calculations of the Brightness-Activity correlation}\label{sect:BYS} 
The magnetic field at the stellar surface affects the measured S-index and photometric flux. In this section we apply the model described in Sect.~\ref{sect:model} to establish the relationship between the observed S-index and the alteration of the Str{\"o}mgren  $(b+y)$/2  flux  caused by the magnetic activity. Hereafter the alteration of the radiative flux is expressed in stellar magnitudes and defined as
%\begin{equation}
%\delta (b+y)/2 \equiv (b+y)_{\rm quiet}/2 \,\, - (b+y)_{\rm total}/2.
\%label{fig:Delta}  
%\end{equation}

\begin{equation}
\frac{\delta (b+y)}{2} \equiv    \frac{(b+y)_{\rm quiet}}{2} \,\, - \frac{(b+y)_{\rm total}}{2}.
\label{fig:Delta}  
\end{equation}

The  $\delta (b+y)/2$ value corresponds to the $F_{\rm active}$ term in Eq.~(\ref{eq:sum})  averaged over the spectral profiles of Str{\"o}mgren  {\it b} and {\it y} filters.  It is positive for a brightness enhancement and negative for a brightness deficit. 

We note that it is not the  $\delta (b+y)/2$ value itself which can be measured, but rather the change of this value relative to some reference point (see Sect.~\ref{sect:SvsF}):
\begin{eqnarray}
\frac{\Delta(b+y)}{2} (t) \equiv   \frac{(b+y)_{\rm total}} {2} (t_0) - \frac{(b+y)_{\rm total}} {2} (t) =  \nonumber \\
 =  \frac{ \delta (b+y)} {2} (t) - \frac{ \delta (b+y)} {2} (t_0), 
\label{eq:Dd}
\end{eqnarray}
where $t_0$ corresponds to the time of the reference measurement. The relative change $\Delta(b+y)/2 $ is defined such that a positive value corresponds to a flux increase  from $t_0$ to $t$.
In this section we will establish the dependence of  $\delta (b+y)/2$ on the S-index.

\subsection{The S-index and photometric brightness as functions of inclination}\label{sect:BYS_i}
Figure~\ref{fig:S_BY} presents the dependences of the photometric brightness and S-index on the stellar inclination for solar,  polar, and homogenous distributions of active regions.  All three curves correspond to different degrees of coverage by faculae and spots, and chosen under the condition that $S_{90}=0.19$, which is close to the maximum  annual value of the solar S-index
over three last cycles (0.188).

One can see that the dependence of $\delta (b+y)/2$ and of S-index on inclination are oppositely directed for  solar and polar distributions of active regions (while, obviously, neither  $\delta (b+y)/2$ nor S-index depend on the inclination for the homogeneous distribution of active regions).  A decrease of the inclination for a star with a solar distribution of active regions leads to their apparent  concentration near the limb  (see Fig.~\ref{fig:ff_solar}). % for such a star observed along its rotational axis, with the active regions covering the near-limb circular belt). 
While negative contrasts of spot umbra and penumbra slightly decrease towards the limb, the positive facular contrast significantly increases (see upper panel of Fig.~\ref{fig:CLV}), so that the effect of the CLV tends to increase the brightness of the star with a solar distribution of  active regions when the observer moves from the equatorial stellar plane to its pole. At the same time the  disc area coverages as seen by such an observer  decrease towards the polar position due to the projection effect (see Eq.~\ref{eq:geom}). For the considered level of activity ($S_{90}^{\rm S}=0.19$) the effect of the CLV  in the Str{\"o}mgren  $b$ and $y$ filters outweighs the effect of projection,  so that the brightness of the star increases with decreasing inclination. The emission in the Ca II K line decreases towards the limb slower than the emission in the nearby continuum (see lower panel of Fig.~\ref{fig:CLV}), so that the CLV effect also tends to amplify the S-index for a star observed along its rotational axis. However, the projection effect now outweighs the CLV effect and the S-index grows slightly  with inclination.

\begin{figure*}
%\resizebox{!}{0.27\vsize}{\includegraphics{HD81809_new}}
%\resizebox{!}{0.27\vsize}{\includegraphics{HD10476_new}}
%\resizebox{!}{0.27\vsize}{\includegraphics{HD82885_new}}
\resizebox{!}{3.0\vsize}{\includegraphics{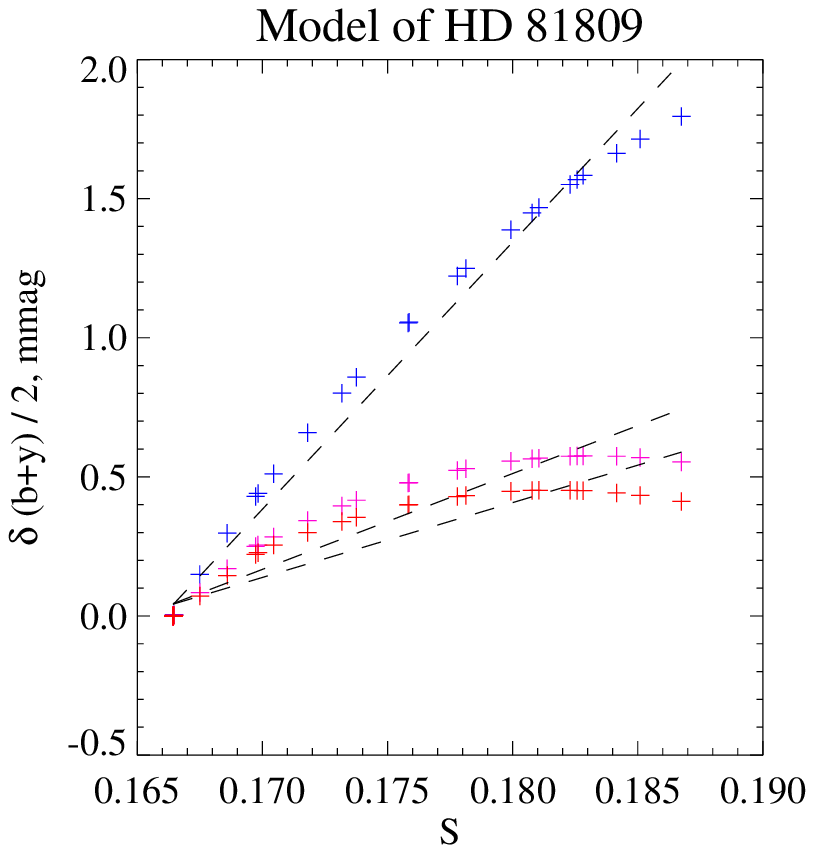}}
\resizebox{!}{3.0\vsize}{\includegraphics{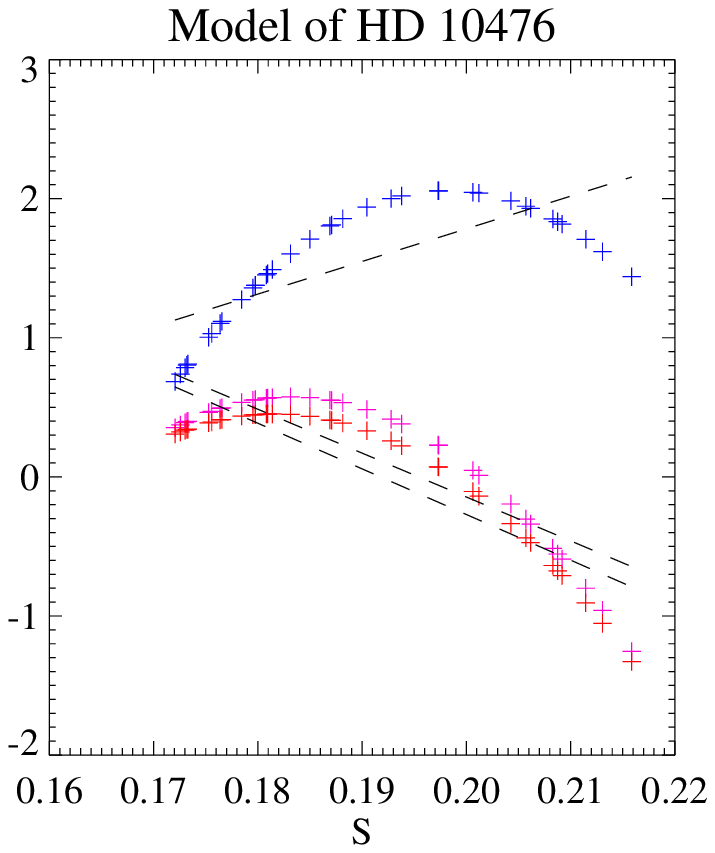}}
\resizebox{!}{3.0\vsize}{\includegraphics{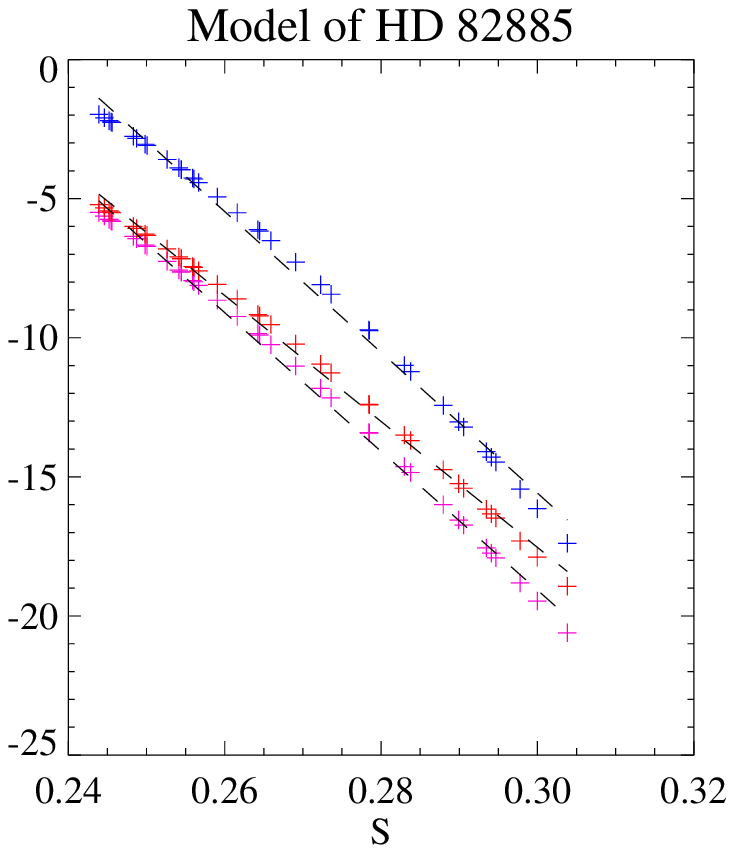}}
\caption{{  Simulated values of the photometric brightness change $\delta (b+y)/2$  as they would be measured  from the equatorial plane (red crosses), at 57$^{\circ}$ inclination (magenta crosses), and from the stellar rotational axis (blue crosses) vs. S-index.  Plotted are values for model stars having the activity level of}  HD 81809 (mean and rms variation values of S are equal to 0.1713 and 0.0080, left panel), of HD 10476 (0.1896 and 0.0136 respectively, middle panel), and of HD 82885 (0.2679 and 0.0186 respectively, right panel). The dashed lines correspond to the linear regressions calculated over the whole $\delta(b+y)/2$ and S-index data sets.}
\label{fig:3HD}
\end{figure*}

The situation is different for  stars with a polar distribution of active regions. For such stars the projection effect amplifies the contribution of active regions when the inclination decreases, so that the $S_i$-index grows significantly with decreasing inclination, reaching higher values at $i=0^{\circ}$  than S$_i^{\rm S}$ at any $i$. This is because at the same total surface coverage by faculae (e.g., solid and dot-dashed curves in Fig.~\ref{fig:S_BY}) the polar distribution at $i=0$ corresponds to active regions more strongly concentrated at large $\mu$ (see Fig.~\ref{fig:ff_polar} in the Online Material).
The contribution of faculae to the photometric brightness, however, diminishes with decreasing inclination. Therefore the contribution of active regions changes from positive  when the star is observed from the equatorial plane to negative  when it is observed from the rotational axis. %We note that if two stars, one with polar and another with solar distribution of the active regions, have the same $S_{90}$-index, then the star with polar distribution has a larger surface area coverage by active regions (as for the same coverage $S_{90}^{\rm P} < S_{90}^{\rm S} $, see Fig.~\ref{fig:FF_S}) and consequently a larger ratio between spot and facular area coverages. 

\subsection{{  Simulated} Brightness-Activity correlation for the Sun}\label{sect:BYS_Sun}
To simulate the dependence of solar brightness on chromospheric activity over the solar cycle we considered the time series of the  annually averaged solar S$_{90}$-index for the period 1977-2011. A similar averaging was done by \cite{lockwoodetal2007}.  The S$_{90}$  values were calculated from the Ca II  Sac Peak data (see Sect.~\ref{sect:FF}).  The mean and rms variation values of  the annual solar S$_{90}$-index over this period are 0.177 and 0.006, respectively. We used this S$_{90} (t)$ time series to calculate the corresponding disc area coverages $A_S (t)$ and $A_F (t)$  (see Sect.~\ref{sect:FF}) and then, employing the  technique described in Sect.~\ref{sect:model}, the $\delta (b+y)_i(t)/2 $ and $S_i(t)$ time series. The self-consistent character of our approach implies that for the inclination $i=90^{\circ}$ the S$_{90} (t)$ values calculated with our approach are equal to the observed S$_{90} (t)$ values used to calculate the disc area coverages and drive the calculations (see Sect.~\ref{sect:model.S}).

Figure~\ref{fig:Sun_angles} {  shows the simulated dependences of the solar brightness on the S-index as they would be seen from three different inclinations}. Independent of the inclination, active regions always increase the annually averaged solar photometric brightness relative to the quiet Sun condition. The $\delta (b+y)/2$ values are larger than zero even for the solar minimum conditions, so that the Sun has not been completely quiet (i.e. free from apparent active regions) for the last three activity cycles. 
Decreasing the inclination simultaneously amplifies the amplitude of the $\delta (b+y)/2$ variability and diminishes the amplitude of the S-index variability (see Fig.~\ref{fig:S_BY}), so that it has a strong effect on the slope of the $\delta (b+y)/2$ vs. S-index  regression. 

As solar activity rises the sunspot area coverage grows faster than the facular area coverage (see Eqs.~\ref{f_spot}--\ref{f_fac}), so that the relative contribution of sunspots to the solar photometric brightness increases with the S-index. This leads to the non-linear profile of the photometric brightness vs. chromospheric activity and causes its saturation (for $i=90^{\circ}$ and $57^{\circ}$) at high activity levels. 
The contribution of faculae increases with decreasing $i$ so that this effect is most pronounced for  the $i=90^{\circ}$ case. For $i=0$ the facular contribution is so strong that it completely dominates over the spot contribution even for the highest solar activity level. As a result, the corresponding dependence of the photometric brightness on activity   is almost linear.

\subsection{Spot- and facula-dominated regimes of variability}\label{sect:SvsF}
If the photometric brightness alteration $\delta (b+y)/2$ is positive, then the facular contribution to stellar brightness (or more specifically to  $F_{\rm active}$, see Eq.~\ref{eq:active}) outweighs the spot contribution. Conversely,  negative values of $\delta (b+y)/2$  imply that the spot contribution outweighs the facular contribution. Therefore, the sign of $\delta (b+y)/2$ provides a clear indication of the relative role of faculae and spots in altering stellar brightness. 

However, the level of irradiance from a star without any magnetic features on the surface  ($(b+y)_{\rm quiet}/2$ in Eq.~\ref{fig:Delta}) is unknown and may even be never reached (as for the Sun, where the calculated  $\delta (b+y)/2$ is always positive). Thus stellar observations do not allow one to measure  $\delta (b+y)/2$. Only the $\Delta (b+y)(t)/2 $ time series,  which is different from the $\delta (b+y)(t)/2 $ time series by a constant term  $\delta (b+y)(t_0)/2 $  (see Eq.~\ref{eq:Dd}), can be measured. Thus, the absolute values of  $\delta (b+y)/2$ are unimportant, so that the curves in Fig.~\ref{fig:Sun_angles} might be arbitrarily shifted up and down. One can see that only the  brightness-activity relationship can be compared to measurements and used to determine the regime of the variability.

Consequently we define the variability of a star as facula-dominated if an increase of the observed S-index leads to an increase  of stellar brightness (i.e. the derivative of the photometric brightness with respect to the observed S-index is positive) and  spot-dominated if it leads to a decrease of the photometric brightness. This definition follows that introduced by \cite{lockwoodetal1992}. The same star can be observed as spot- or faculae-dominated at different levels of activity. For example, the Sun observed at  $i=90^{\circ}$ is faculae-dominated at low and intermediate activity, but it can become spot-dominated for a very short time around the maximum of activity  (see Fig.~\ref{fig:Sun_angles}).

\subsection{Brightness-Activity correlation for Sun-like stars}\label{sect:BYS_stars}
To {  model} the brightness-activity relationships for Sun-like stars we linearly scale the annual values of the solar S-index for the 1977-2011 period (which covers solar cycles 21--23 and the beginning of cycle 24) to reproduce the measured stellar mean and rms variation values of the S-index on various stars, where the mean and rms variation are taken over the whole period that the star is observed.  We note that in reality stellar cycles need not resemble the solar cycle and thus our approach is only an approximation.

Unlike the solar case, the observed stellar S-index corresponds to an arbitrary inclination $i$. Therefore, we first assume that the star has a given inclination $i$ and convert the observed mean and rms variation values of S$_i$ to the mean and rms variation values of S$_{90}$, using the technique described in Sect.~\ref{sect:model}. Then we construct the S$_{90}$(t) time series as described in the above paragraph and use it 
 to calculate  $\delta(b+y)_i/2$  and S$_i$ (see Sect.~\ref{sect:model}). The self-consistent character of our approach guarantees that the S$_i$ values used to feed the model are identical to the values calculated in this final step. This procedure is repeated for a number of $i$ values.

In  Fig.~\ref{fig:3HD} we plot the modelled brightness-activity dependences for HD 81809 which has a level of activity slightly smaller than the solar value, for HD 10476, which is slightly more active than the Sun, and for HD 82885, which is significantly more active than the Sun. The mean and rms variation values of the observed S-index for these stars have been provided by Lockwood (2013, personal communication). The calculations were performed assuming a solar distribution of active regions. We note that unlike Fig.~\ref{fig:Sun_angles}, where we considered the hypothetical case of the Sun observed out-of-ecliptic, 
the amplitude of the S-index (or rather S$_i$-index) variability does not depend on the assumed inclination and always corresponds to the observed amplitude {  which is used to calculate the S$_{90}$(t) time series}.

The $\delta (b+y)/2$ values of HD 81809 are always positive, independent of the inclination. This is not surprising, given its similar activity level to the Sun. The star is observed as faculae-dominated most of the time. However, for  $i=57^{\circ}$ and $90^{\circ}$ starting from $S \simeq 0.18$, the dependence of the brightness on activity is saturated so that the further increase in activity has almost no effect on brightness.  We note that the minimum value of the observed S-index for  HD 81809 ($\approx 0.166$) is smaller than the solar minimum  value  ($\approx 0.169$)  so that, unlike the solar case,  $\delta (b+y)/2$ reaches zero. A relatively robust measure of whether the variability of a star is faculae- or spot- dominated {\it over the period of observations} is the slope of the regression of brightness on activity calculated employing the whole  $\delta (b+y)/2$ and S-index time series (dashed lines in Fig.~\ref{fig:3HD}).

The variability of HD 10476 is faculae-dominated for the low values of the observed S-index and spot-dominated for the high values at all $i$. The overall correlation is direct for $i=0^{\circ}$, and inverse for  $i=57^{\circ}$ and  $i=90^{\circ}$. Thus HD 10476 presents a curious case of a star whose regime of variability {\it over the period of observations } depends on the inclination. The activity of HD 82885 is so high that the facular contribution is negligibly small, and its variability is always spot-dominated. %We note that the real stellar $S$-index time series may be different from the scaled solar time series considered here. Thus the dependences plotted in Fig.~\ref{fig:3HD} are correct only in an approximate sense.

\begin{figure}
\resizebox{\hsize}{!}{\includegraphics{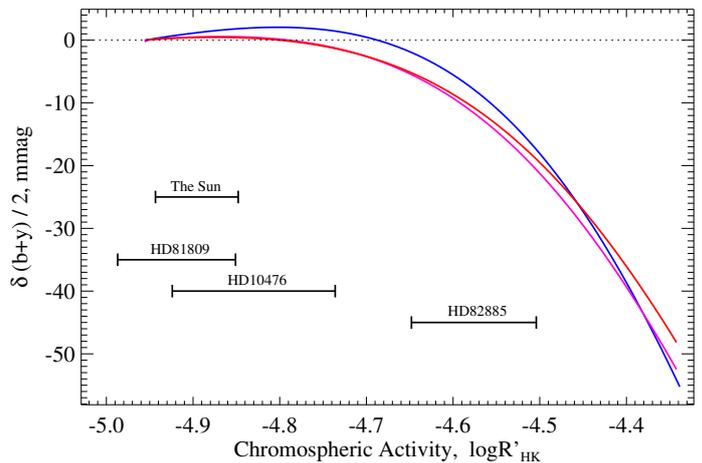}}
\caption{{  Simulated values of the photometric brightness change $\delta (b+y)/2$  as they would be measured  from the equatorial plane (red curve), at 57$^{\circ}$ inclination (magenta curve), and from the stellar rotational axis (blue curve) vs. S-index.  Unlike Fig.~\ref{fig:3HD} the dependences are calculated over a broad range of stellar magnetic activity.  The range bars designate the observed ranges of chromospheric activity variations of the Sun, HD 81809, HD 10476, and HD 82885. }}
\label{fig:HD_large}
\end{figure}

The discussion above was limited to the S-index and photometric time series of specific stars.  The approach presented in Sect.~\ref{sect:model} also allows one to {  model} the dependence of the photometric brightness on the chromospheric activity over a broad range of S values. These dependences are plotted in Fig.~\ref{fig:HD_large} for three values of the inclination and assuming a solar distribution of active regions. The chromospheric activity is characterised via the ${\rm log R'_{ HK}}$ parameter, which is a function of the S-index and $B-V$ colour \citep[see][for a detailed discussion]{radicketal1998}. Hereafter, we will use the solar value of the color ($B-V=0.65$) to convert the S-index to ${\rm log R'_{ HK}}$ {(e.g. the range bars in Fig.~\ref{fig:HD_large} are also calculated assuming solar $B-V$ value)}. This is equivalent to keeping the effective temperature of the hypothetical Sun unchanged, while increasing its activity level.

Figure~\ref{fig:HD_large} reveals that according to our model low values of activity correspond to faculae-dominated variability, while high values of activity correspond to spot-dominated variability. For a solar distribution of active regions the lower the inclination of the star, the stronger the amplification of the facular contribution by the CLV effect.  Therefore the threshold activity which corresponds to the transition between the spot- and facuale-dominated regimes is higher for lower inclination values (blue vs. red curves in Fig.~\ref{fig:HD_large}).
Comparing Fig.~\ref{fig:3HD} with Fig.~\ref{fig:HD_large} one can see that the  individual stellar Brightness-Activity relationships  presented in the three panels of Fig.~\ref{fig:3HD} are different parts of the general dependence.  

{  Up to this point we have established the connection between stellar brightness and activity and discussed different regimes of variability, refraining from a comparison with actual data. In the next Section we show how our approach can be employed to model the general patterns of stellar variability obtained by the Lowell and Fairborn programs and published by \cite{radicketal1998, lockwoodetal2007}. }

\section{Variability patterns:  comparison of model with observations}\label{sect:pat}
%In Sect.~\ref{sect:BYS} we discussed the individual stellar brightness-activity relationships. In this section, we will employ our model to reproduce the general patterns of stellar variability obtained by the Lowell and Fairborn programs (see Sect.~\ref{sect:intro}). 

{  For every star from the Lowell and Fairborn programs, the two time series, $(b+y)(t)/2$   and  S(t), are  simultaneously available.} In the literature the amplitude of the photometric variability is usually characterised via $\log({\rm rms} (b+y)/2) $ (expressed in stellar magnitudes). The connection between photometric brightness and chromospheric activity is expressed via the slope of the regression of photometric brightness on the observed S-index $\Delta[(b+y)/2]/\Delta S$ (see the dashed lines in  Fig.~\ref{fig:3HD}). Usually the whole available time series are used to calculate these values, so that the variability of every star is described by the two numbers, $\log({\rm rms} (b+y)/2) $ and $\Delta[(b+y)/2]/\Delta S$. 
Positive $\Delta[(b+y)/2]/\Delta S$ values correspond to the  faculae-dominated regime of variability {\it over the period of observations}. The restriction to the particular period of variability is important since a given star may appear faculae or spot dominated at different times, so that having a long time series is important. Negative  $\Delta[(b+y)/2]/\Delta S$ values correspond to the spot-dominated regime of variability.

%In Sect.~\ref{sect:BYS_stars} we showed how the time-series of the observed chromospheric activity $S(t)$ might be used to calculate the time series of the brightness alteration caused by active regions $\delta(b+y)(t)/2$.
% We note that the  $ (b+y)(t)/2   $ and $\delta(b+y)(t)/2$ time series have the same rms variabilities and the coefficient of the regression of photometric brightness on the S-index is not affected by the constant difference between $\delta(b+y)(t)/2$ and $\Delta(b+y)(t)/2$ time series (see Sects.~\ref{sect:SvsF} and middle panel of Fig.~\ref{fig:3HD}). 
%Thus our model allows us to calculate $\log({\rm rms} (b+y)/2) $ and $\Delta[(b+y)/2]/\Delta S$ values from the observed S(t) time series for any arbitrary inclination and latitudinal distribution of active regions.

The goal of this section is to establish the dependences of $\log({\rm rms} (b+y)/2) $ and $\Delta[(b+y)/2]/\Delta S$ values on the mean level of  chromospheric activity and compare them to the {  observed} dependences. 
\cite{lockwoodetal2007} found that the variability of the S-index increases from less active to more active stars and established an empirical linear relationship between the variability of  chromospheric activity ${\rm log (rms \, R'_{ HK}})$ and mean activity ${\rm log R'_{ HK}}$ (see their Fig. 6). This relationship may be used to connect the rms variation and the mean values of the S-index (assuming  the solar value for the colour B-V which is needed to convert ${\rm log R'_{ HK}}$  into the S-index). The resulting relationship between the rms variation and the mean values of S-index is established for a group of stars with random distribution of inclinations and, thus, connects values which are averaged over all possible inclinations (with a weighting factor $\sin i$, which is proportional to the probability that a random star has an inclination $i$). Using the technique described in Sect.~\ref{sect:model.S} we transformed this relationship into the relationship which connects the rms variation and the mean values of the S-index expected when observing in the equatorial plane. 

%We note that the transformation from the real observed relationship to the relationship expected when observing in the equatorial plane depends on the latitudinal distribution of active regions (see Fig.~\ref{fig:S_BY}) . Following the approach of Sect.~\ref{sect:model}~and~\ref{sect:BYS} we again considered the two opposite cases of solar and polar latitudinal distributions. In principle, one might expect that the observed relationship between the rms variation and the mean values of S is affected by the latitudinal distribution of active regions and, thus, consists of several branches. However,  the identification of these branches is not straightforward because of the limited amount of stars used to establish the relationship. Thus for simplicity we assumed that the relationship between observed  rms variation and the mean values of the S-index does not depend on the latitudinal distribution of active regions. This implies that the relationship expected when observing in the equatorial plane depends on the latitudinal distribution of active regions.

We consider a set  of mean  chromospheric activity $<{\rm S}_{90}>$ values with minimal value $0.167$ (which is the minimum annual value of the solar S-index) and maximal value $0.5$. For each  $<{\rm S}_{90}>$ value  we compute corresponding values of the rms variation assuming different distributions of active regions and then generate the S-index time series by scaling the solar S-index time series. Next, we employ these time series to calculate $\log({\rm rms} (b+y)/2) $, $\Delta[(b+y)/2]/\Delta S$, and $<S>$ values  for several inclinations and active regions distributions. %We note that $<S>$, which is the value observed assuming some specific inclination and latitudinal distribution of active regions, might be different from  $<{\rm S}_{90}>$, which  corresponds to the observations from equatorial plane and solar distribution of active regions (see Sects.~\ref{sect:model.S}--\ref{sect:model.polar}). 
The dependence of $\log({\rm rms} (b+y)/2) $ and of $\Delta[(b+y)/2]/\Delta S$ on mean chromospheric activity ${\rm log R'_{ HK}}$ is presented in Sect.~\ref{sect:pat.Ph} and Sect.~\ref{sect:pat.tr}, respectively.

\subsection{Photometric variability versus mean activity}\label{sect:pat.Ph}
\begin{figure}
\resizebox{\hsize}{!}{\includegraphics{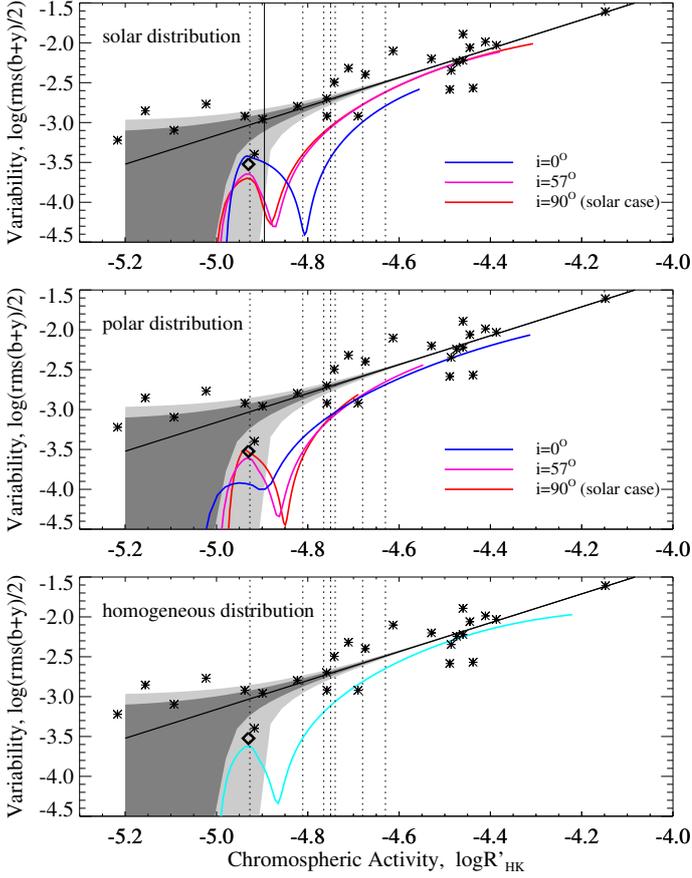}}
\caption{{  Comparison of the observed and modelled photometric variability. The modelled values of the photometric variability are plotted vs. mean chromospheric activity for model stars with solar (upper panel), polar (middle panel), and homogeneous (lower panel) latitudinal distributions of active regions.} The {  asterisks} and the black lines indicate stars with observed variability and the regression from \cite{lockwoodetal2007}. {  The dark (light) shaded areas indicate estimated $1\sigma$ ($2\sigma$) uncertainty in the \cite{lockwoodetal2007} data.} The diamond indicates 18 Scorpii (HD 146233) from \cite{halletal2009}.  Coloured curves result from our calculations for three values of the stellar inclination: $90^{\circ}$ (red curve), $57^{\circ}$ (magenta curve), $0^{\circ}$ (blue curve). The dotted vertical lines denote the mean level of chromospheric activity of stars with unconfirmed variability (only stars with  ${\rm log R'_{ HK}}>-5$ are shown). The solid vertical line in the top panel shows the mean level of solar chromospheric activity.}
\label{fig:ampl}
\end{figure}

In Fig.~\ref{fig:ampl} we plot the stellar photometric variability $\log({\rm rms} (b+y)/2) $   vs. mean chromospheric activity ${\rm log R'_{ HK}}$, comparing the data of   Lockwood et al. (2007, stars and regression line) with our calculations (coloured curves). {  The shaded areas around the regression lines define the uncertainty in stellar variability as a function of chromospheric activity. This has been estimated by accounting for the uncertainty in the variance measurements for the target and the comparison stars. Most of the program stars in the \cite{lockwoodetal2007} sample have two suitable comparison stars. \cite{lockwoodetal2007} defined the variability of a program star as  the observed variance $\sigma_{1,23}^2$  (relative to the mean of the two comparison stars) minus one half of the observed variance of the comparison star pair $\sigma_{2,3}^2$. \cite{radicketal1998} adopted the uncertainty of the variance $\epsilon= 0.0006$ mag. This uncertainty is the same for Lowell and Fairborn observations and does not depend on the stellar magnitude \citep{lockwoodetal2007}. We use this value to estimate $1\sigma$ and $2\sigma$ uncertainties in the observed photometric variability, assuming that both $\sigma_{1,23}$ and  $\sigma_{2,3}$ values are known with the uncertainty $\epsilon$. We note that it is not easy to properly take into account the variability of the comparison stars, especially for the low activity program stars with small photometric variability (actually several comparison stars appeared to be more variable than the program stars). Thus our simple calculations provide a rather low estimate of the uncertainty. }

{While our calculations are in  a reasonably good agreement with the empirical regression for the high activity stars, the variabilities of the low activity stars  given by our model are significantly below the observations. Interestingly, the variability of 18 Scorpii, which is arguably the closest bright solar analog \citep{halletal2007a, analogs,petitetal2008} is correctly reproduced by our model (see Fig.~\ref{fig:ampl}). 

Models and measurements indicate that the solar irradiance variability on the 11-year cycle time scale is smaller than given by the  \cite{lockwoodetal2007} empirical regression of the stellar variability vs. activity  \citep[see e.g.][and references therein]{knaacketal2001,analysis,shapiroetal2013_stars}. As our model is based on the SATIRE, that accurately reproduces measured TSI variability, it is no surprise that it also leads to the same deviation.

One of the reasons for the disagreement between the values of the photometric variability of low activity stars given by our model and obtained with the  \cite{lockwoodetal2007}  empirical regression comes from the uncertainties in the observational data. The  \cite{lockwoodetal2007}  regression may be affected by the short sample length (less than 20 years), small size (3 dozen stars) and possibly a selection effect \citep[see e.g.][]{halletal2009} so that the Sun-like stars with small variabilities are not included in the sample. Thus it would be important to repeat the comparison when more stellar data become available and the low end of the stellar activity sequence (which is the most difficult to study observationally) is better constrained.}

The conspicuous feature of our {  calculations} is a strong decrease of the photometric variability near  ${\rm log R'_{ HK}}=-4.8$ (hereafter called the variability gap), {  which is absent in \cite{lockwoodetal2007} data. In our calculations the variability gap} is caused by the compensation of the spot and facular contributions to the variability, and consequently the transition from the faculae-dominated (on the left side of the gap) to spot-dominated (on the right side of the gap)  regime of variability. %Indeed, Fig.~\ref{fig:HD_large} shows that the alteration of the photometric brightness caused by active regions, $\delta (b+y)/2$, has a maximum around ${\rm log R'_{ HK}}=-4.8$  and, thus, any change in chromospheric activity around this ${\rm log R'_{ HK}}$ value only marginally affects the photometric brightness. 
The level of activity (traced by ${\rm log R'_{ HK}}$), at which the variability gap occurs depends on the inclination and latitudinal distribution of active regions.
For a solar distribution the decrease of the inclination shifts active regions towards the limb (see Fig.~\ref{fig:ff_solar}) and amplifies the facular contribution, shifting the variability gap to larger ${\rm log R'_{ HK}}$ values. The polar distribution displays the  opposite behavior (see Fig.~\ref{fig:ff_polar}) and the decrease of the inclination shifts the variability gap to lower ${\rm log R'_{ HK}}$ values. 

%We note, however, that the direct comparison of the cases with different inclinations and distributions of the active regions is not straightforward, as the same  ${\rm log R'_{ HK}}$ values correspond to a different coverage by active regions (see Sect.~\ref{S_proxy}).

Interestingly, the chromospheric activity levels of the stars, whose variability \cite{lockwoodetal2007} were not able to confirm, cluster around the variability gap predicted by our model (see Fig.~\ref{fig:ampl}). There are, however, also stars with confirmed variability located in the variability gap. {  The variability of such stars cannot be reproduced by our model. We note that these are exactly the stars which cause the deviation between  the \cite{lockwoodetal2007}  regression and our calculations (here we ignore the four stars with ${\rm log R'_{ HK}}<-5$ since their variability is quite uncertain and our model cannot reproduce the variability of stars with such a small magnetic activity, see Sect.~\ref{sect:pat.tr}). The absence of the gap in the observational data and consequently the deviation between the \cite{lockwoodetal2007}  empirical regression and our model can be the result of one or more of the following:

%1. The position of the variability gap depends on the latitudinal distribution of active regions and the inclination (see Fig.~\ref{fig:ampl}).

1. The dependences of the spot and facular disc coverage on the S-index  (Eqs.~\ref{f_spot}--\ref{f_fac}) employed in our model are rather approximate and may also vary from star to star.
This may have a strong effect on the variability of stars around the gap. Indeed, 
%the variability of the stars on the left side of the gap is entirely dominated by faculae, while the variability of the stars on the right side of the gap is dominated by spots. At the same time, 
the variability of such stars is determined by the balance between spot and facular contributions.  A small change of the ratio between spot and facular surface coverages (as well as between spot and facular brightness contrasts, see point 3) may break this delicate balance and thus strongly affect the variability of the stars around the gap (see also Appendix B).  In contrast, such a change only marginally affects the variability of stars far from the gap.   

2. Our model only accounts for the photometric variability on the activity time scale. The measured stellar variability may be affected by the long-term variability and short-term variability on the time-scale of stellar rotation, which may be not completely eliminated by the annual averaging performed by \cite{lockwoodetal2007}. Since, unlike the case of the variability on the activity time scale, we do not expect any compensation effect in the rotational and long-term variabilities, they may significantly contribute to the total variability around the gap \citep[see ][for a more detailed discussion and estimations]{shapiroetal2013_stars}. 

3. The position of the gap very likely depends on B-V since the facular and sunspot contrasts are expected to depend on the effective temperature of the star. It is possible that while the Sun is located in the variability gap, other stars with similar levels of magnetic activity but higher photometric variabilities are located outside of the gap.

4. The stellar variabilities may be affected by a not yet identified physical mechanism which is not taken into account by our simple extrapolation from the Sun.}

%It will also be important to check how the model curves depend on B-V, since that will allow us to check if the low variability of the Sun is due to its lying in the variability gap, while other stars showing higher variability.

\subsection{Faculae- and spot-dominated stars}\label{sect:pat.tr}
One quantity which allows distinguishing easily between stars whose photometric variability is dominated by faculae and those with spot-dominated photometric variability is the sign of the change in brightness {with changing chromospheric activity.} %The relative amplitudes of the two types of variability give a measure of how strong the dominance of faculae or spot is.
\cite{lockwoodetal1992} introduced the slope of the regression to photometric brightness vs. S-index  $\Delta[(b+y)/2]/\Delta S$  as a measure of faculae- or spot-dominance. %Thus each one year average of their photometric brightness measurements of a given star is plotted vs. the S-index averaged over the same period of time. The slope of the regression  to all measurements of the star, $\Delta[(b+y)/2]/\Delta S$, is then plotted vs. the average activity level of the star, given by  ${\rm log R'_{ HK}}$.
The zero value of   $\Delta[(b+y)/2]/\Delta S$ corresponds to the threshold between faculae- and spot-dominated regimes of photometric variability.

\begin{figure}
\resizebox{\hsize}{!}{\includegraphics{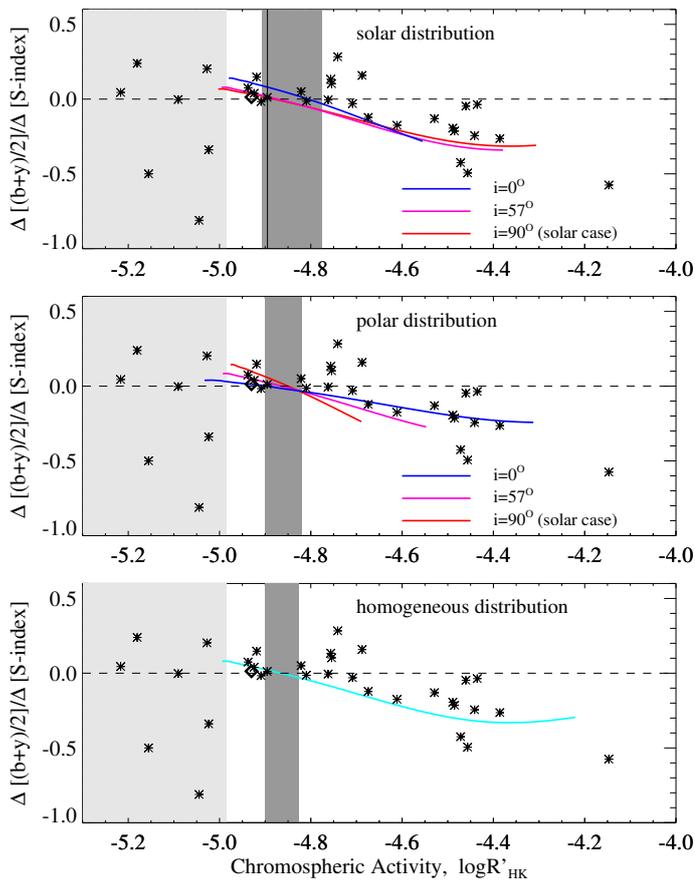}}
\caption{Slope of the regression to the dependence of photometric brightness variation on HK emission variation, plotted vs. mean chromospheric activity ${\rm log R'_{ HK}}$ for stars with solar (upper panel), polar (middle panel), and homogeneous (lower panel) distributions of active regions. The {  asterisks}  indicate the observed values for the stellar sample of \cite{lockwoodetal2007}.  {  The diamond indicates 18 Scorpii (HD 146233) from \cite{halletal2009}. The light shaded areas represent the activity levels for which photometric variability, according to the activity-variability regression from \cite{lockwoodetal2007}, is smaller than the $1\sigma$ uncertainty and thus $\Delta[(b+y)/2]/\Delta S$ values cannot be reliably defined}. Coloured curves are the output of our model calculated for three values of the stellar inclination: $90^{\circ}$ (red curve), $57^{\circ}$ (magenta curve), and $0^{\circ}$ (blue curve).
The dashed lines separate the facula-dominated (positive $\Delta[(b+y)/2]/\Delta S$)  from the spot-dominated (negative $\Delta[(b+y)/2]/\Delta S$)  variability.  The solid vertical line on the upper panel denotes the mean level of solar chromospheric activity. The dark shaded bands indicate the range of the chromospheric activities for which {  according to our model} the stars can be observed as either faculae or as spot-dominated, depending on the period of time over which they are observed (see text for details).}
\label{fig:regr}
\end{figure}

{  In  Fig.~\ref{fig:regr}  we plot  $\Delta[(b+y)/2]/\Delta S$ values given by \cite{lockwoodetal2007} and similarly computed with our model. As in Fig.~\ref{fig:ampl} the three panels differ only in the spatial distribution of active regions on the stellar surface assumed for the model (the observed data, asterisks, are the same in all panels). Most of the observed  stars are located in between the synthetic curves,  so our results are in good agreement with the data of  \cite{lockwoodetal2007}.  For the spot-dominated stars our model reproduces the increase of photometric variability relative to chromospheric variability with increasing activity level.

{  Interestingly, three stars in Fig.~\ref{fig:regr} appear to be spot-dominated despite the low level of their mean chromospheric activity. Our simple extrapolation from the Sun cannot reproduce such low values of the mean chromospheric activity (one would have to adjust the value of $S^Q$ for this; see Eq.~\ref{S_def}), which implies that the temperature structures of the quiet and magnetic regions of these stars are different from the respective solar temperature structures.
At the same time these stars are located in the light grey shaded region in Fig.~\ref{fig:regr}, which implies that their photometric variabilities are below the uncertainty level and consequently $\Delta[(b+y)/2]/\Delta S$ values are quite uncertain. For example, one of these stars, HD 14376, was also observed by \cite{halletal2009} who found no activity-brightness correlation, instead of the inverse activity-brightness correlation found by \cite{lockwoodetal2007}. }

If observed stellar photometric brightness is affected by a systematic trend or noise  (which may be stellar and/or instrumental in nature, see Sect.~\ref{sect:pat.Ph}), then it will 
have a stronger effect on the measured photometric variability than on the $\Delta[(b+y)/2]/\Delta S$ values. This may explain why the 
observed data and our model are in a better agreement in Fig.~\ref{fig:regr} than in Fig.~\ref{fig:ampl}}.
 
 \cite{radicketal1998} suggested that the term ``Sun-like''  might be better applicable to the older and low activity stars with faculae-dominated variability than for the stars with spot-dominated variability. Our results indicate that the qualitative behaviour of the more active Sun-like stars can be reproduced using the Sun as a model star simply by extrapolation. Based on this we can conclude that even the younger stars with spot-dominated variability might be considered as ``Sun-like'' stars  in the sense that their variability is defined by the same physical processes as the solar one.

%These slopes determine the sensitivity of the stellar photometric brightness to the change of the activity (as traced by the S-index). The stars with positive slopes are deemed as facular dominated, while the stars with negative slopes are considered as spot dominated. We note that, strictly speaking, this definition is not very accurate as the sign of the $\Delta (b+y)/2$ value changes at different activity than the sign of the  $\Delta (b+y)/2$ derivative (see Fig.~\ref{fig:HD_large}). So there is a range of activities for which the stellar brightness is enhanced by active regions ($\Delta (b+y)/2$ is positive) and the star is facular dominated but  deemed as spot dominated. We will hereafter stick to the definition via the regression slopes  as only the variably of $\Delta (b+y)/2$ and not its  absolute value may be measured.

We note that the Rossby number which defines the efficiency of the stellar dynamo \citep{noyesetal1984} is supposed to increase with stellar activity. At the same time the latitudes of the magnetic field emergence depend on the {\it magnetic} Rossby number \citep{polar1}. The actual latitudes of emergence of the field also depend on a number of further parameters, including the depth of the convection zone  \citep[i.e. on B-V and on the evolutionary state of the star, see][]{polar2, granzer2002}. In addition, meridional flows may move magnetic features towards the poles even after emergence \citep{SchrijverandTitle2001}. Therefore, while, in general, one might expect that  the solar distribution of active regions is more representative for  low activity stars and  the polar distribution for  high activity stars, there are likely to be deviations from this straightforward rule and the stars with the same  ${\rm log R'_{ HK}}$ may have different latitudinal distributions of active regions. The comparison of the slopes calculated with {  different} distributions allows one to estimate the scatter in the observed values which might be attributed to different  latitudinal distributions of active regions.

%We note that the significant scatter below  ${\rm log R'_{ HK}}=-5$ may be attributed to the poorly known photometrical variability of these stars and most probably is not real \citep{lockwoodetal2007}. 

Our calculations indicate that the transition from  the facula- to the spot-dominated regime occurs somewhere between   ${\rm log R'_{ HK}}=-4.9$ and  ${\rm log R'_{ HK}}=-4.7$, which agrees well with the \cite{lockwoodetal2007} and \cite{halletal2009} observations. The activity level of the transition depends also  
 on the inclination and latitudinal distribution of active regions (and probably on B-V, which is however outside of the scope of this study), so that the regimes of  variability are not sharply defined,  {  which is in line with the observations of  \cite{halletal2009}}. While according to our calculations the solar variability is faculae-dominated \citep[though see][]{harderetal2009, premingeretal2011}, the Sun is located very close to the threshold between the regimes, so that the stars which are a bit more active than the Sun might be observed as spot-dominated.

The mean and rms variation values of the chromospheric activity do not uniquely define the S-index time series. This is an extra source of the scatter in
Fig.~\ref{fig:regr} because the $\Delta[(b+y)/2]/\Delta S$ values might be affected by the specific form of the  S-index time variability. We note that the  theoretical $\Delta[(b+y)/2]/\Delta S$ values   presented here were calculated employing the scaled solar S-index time series (see Sect.\ref{sect:BYS_stars}), which is only an approximation. Additionally the stellar observed time series may be too short to reveal a true regime of the variability. For example if the Sun is observed for a short period of time around its activity maximum from its equatorial plane it may be falsely identified as spot-dominated (see Fig.~\ref{fig:Sun_angles}). This effect may explain some part of the scatter in the observed slopes  in Fig.~\ref{fig:regr} and points to the importance of having long time-series to accurately clarify such stars. {  The effect of the possible deviations in the stellar disc area coverages by active regions from the dependences established in Sect.~\ref{sect:FF} is discussed in Appendixes~\ref{app:dev}~and~\ref{sect:ratio}. }

%Another effect is discussed in Appenidx~\ref{sect:ratio}.

If the varying stellar activity crosses the threshold between the facular and spot-dominated variability regimes  then the star can appear to be facula-dominated over some periods of time (when the activity is below the  threshold), while it will be observed to be spot-dominated over other intervals of time  (when the activity is above the threshold).  In Fig.~\ref{fig:regr} we shaded the  ${\rm log R'_{ HK}}$ ranges which correspond to such regimes of variability for both latitudinal  distributions of active regions.
As expected the Sun lies within the shaded area. Another interesting example of a ``cross over'' star is HD 140538 ($\psi$ Ser), which demonstrates both direct and inverse activity-brightness correlations of timescales of four years  \citep{halletal2009}. We note, however,  that  the complete picture of its variability might be also affected by many other factors, including the change of the latitudinal distribution of active regions,  change of faculae to spot coverage ratio, and observational noise.

%\begin{figure}
%\resizebox{\hsize}{!}{\includegraphics{var_corr.eps}}
%\caption{Activity-brightness correlation as a function of mean chromospheric activity for stars with polar (dashed curves) and solar (solid curves)  distributions of active regions. The calculations are shown for for three values of the stellar inclination: $90^{\circ}$ (red curve), $57^{\circ}$ (magenta curve), $0^{\circ}$ (blue curve).}
%\label{fig:st_corr}
%\end{figure}

% Here we start

% Here we finish

\section{Conclusions}\label{sect:conc}
A long-standing puzzle in the study of stellar activity has been the observation that whereas on one hand stars with a relatively low level of activity, become photometrically brighter (averaged over a year) as their activity level increases, on the other hand more active stars display the opposite behavior, becoming darker with rising activity level. We  reproduce this phenomenon based on the assumption that the solar paradigm is also valid for more active stars, i.e. we can qualitatively describe stellar behavior by extrapolating from solar activity and brightness variations.

We have expanded a simplified version of the SATIRE model of solar variability to Sun-like stars with different levels of magnetic activity.  Our model attributes the variability of Sun-like stars to the spatial inhomogeneities (i.e. bright faculae and dark spots) on their surfaces, caused by the presence of  a magnetic field at the solar/stellar surface. The key ingredient to the extrapolation is the finding that the area coverage by sunspots increases more rapidly with chromospheric activity than the coverage by faculae \citep[see also][]{foukal1998}, although only the latter contribute strongly to the Ca II H and K line core emission, which determines the S-index.

%It allowed us  to calculate the photometric variabilities of Sun-like stars as seen in Str{\"o}mgren filters (b+y)/2 and compare them with available ground-based measurements.

We have employed our model to study the dependences of the stellar photometric variability on the observed mean level of magnetic activity (as traced by the Ca II S-index), stellar inclination (i.e. the angle between the direction to the observer and stellar rotational axis) and, at a simple level, also on the latitudinal distribution of active regions on the stellar surface.

We have found that our simple extrapolation from the Sun to higher activity stars reproduces the observed trend that while the variability of Sun-like stars with low magnetic activity is  dominated by faculae (i.e. photometric brightness and magnetic activity are positively correlated), the variability of  Sun-like stars with high magnetic activity is spot-dominated (photometric brightness and magnetic activity are negatively correlated). The switch between the regimes happens around magnetic activity levels slightly larger than solar and depends on the inclination and latitudinal distribution of active regions. The slopes of the activity-brightness correlation, calculated with our model, are  in good agreement with the \cite{lockwoodetal2007} data.
Our calculations indicate that the variability of a star with a solar level of magnetic activity is faculae-dominated, independently of the inclination and latitudinal distribution of active regions on the surface, although a too short time series of solar activity, if caught at the right phase may lead to the opposite conclusion.

The measured photometric variabilities of the high activity stars agree with our calculations reasonably well. However, the variabilities of the low activity stars  calculated with our model, are smaller than given by the empirical correlation between the stellar photometric variability and mean chromospheric activity level. {  This is probably caused by the limitations of our simple approach and by the uncertainties in the stellar measurements.}

%It is presently unclear whether this deviation is caused by the uncertainties in the stellar measurements or by the limitations of our approach. % Interestingly, according to our model the Sun is also less variable  than indicated by the empirical relationship. 

{  The general success of the model in reproducing the basic qualitative behavior of spot-dominated stars is an indication that the photometric variability of more active stars has the same fundamental causes as the Sun's.} Up until now physics-based models of irradiance variability were solely applied to the solar case. Consequently, they could only be validated and constrained by solar data, which represent a single point in a wide parameter space of the possible magnetic activities, inclinations, latitudinal  distribution of active regions, etc. The approach presented in this paper allows constraining the model over a much wider parameter space, and, thus, along with interpreting stellar data, it helps to better understand the mechanisms of  solar variability.

{  As a next step we plan to apply an extension of this model to study stellar variability on rotational time scales, as observed by the COROT \citep{COROT} and Kepler \citep{KEPLER} missions and in future to be measured by the PLATO mission \citep{PLATO}.}

%Even the ratio of starspot to stellar facular coverage can be obtained by extrapolating from the Sun. We construe this as a remarkable success of the solar paradigm of stellar magnetic activity.

%We have demonstrated that the compensation between facular and spot contributions to the photometric brightness results in a significant decrease of the  photometric variability  for stars slightly more active than the Sun. % This decrease is hinted in the \cite{lockwoodetal2007} and \cite{halletal2009} data.

%??? Should I also mention that S-index is a bad proxy?

\begin{acknowledgements}
{  We thank Wes Lockwood for useful discussions and for helping us to estimate the uncertainty in stellar data and the referee Gibor Basri for constructive criticism and useful advise.}
The research leading to this paper was supported by the Swiss National Science Foundation under grant CRSI122-130642 (FUPSOL). 
It also got a financial support from the COST Action ES1005 TOSCA (http://www.tosca-cost.eu) and from the BK21 plus program through the National Research Foundation (NRF) funded by the Ministry of Education of Korea. WTB and YCU acknowledge support through STFC grant ST/I001972/1.  AIS and WKS acknowledge support from SNF grant 100020\_140573.
\end{acknowledgements}

\bibliographystyle{aa}
%\bibliography{shapiro}

\begin{thebibliography}{73}
\expandafter\ifx\csname natexlab\endcsname\relax\def\natexlab#1{#1}\fi

\bibitem[{{Baglin} {et~al.}(2006){Baglin}, {Auvergne}, {Boisnard}, {Lam-Trong},
  {Barge}, {Catala}, {Deleuil}, {Michel}, \& {Weiss}}]{COROT}
{Baglin}, A., {Auvergne}, M., {Boisnard}, L., {et~al.} 2006, in COSPAR Meeting,
  Vol.~36, 36th COSPAR Scientific Assembly, 3749

\bibitem[{{Baliunas} {et~al.}(1998){Baliunas}, {Donahue}, {Soon}, \&
  {Henry}}]{baliunasetal1998}
{Baliunas}, S.~L., {Donahue}, R.~A., {Soon}, W., \& {Henry}, G.~W. 1998, in
  Astronomical Society of the Pacific Conference Series, Vol. 154, Cool Stars,
  Stellar Systems, and the Sun, ed. R.~A. {Donahue} \& J.~A. {Bookbinder}, 153

\bibitem[{{Baliunas} {et~al.}(1995){Baliunas}, {Donahue}, {Soon}, {Horne},
  {Frazer}, {Woodard-Eklund}, {Bradford}, {Rao}, {Wilson}, {Zhang}, {Bennett},
  {Briggs}, {Carroll}, {Duncan}, {Figueroa}, {Lanning}, {Misch}, {Mueller},
  {Noyes}, {Poppe}, {Porter}, {Robinson}, {Russell}, {Shelton}, {Soyumer},
  {Vaughan}, \& {Whitney}}]{HK}
{Baliunas}, S.~L., {Donahue}, R.~A., {Soon}, W.~H., {et~al.} 1995, \apj, 438,
  269

\bibitem[{{Ball} {et~al.}(2013){Ball}, {Krivova}, {Unruh}, {Haigh}, \&
  {Solanki}}]{ball_climate}
{Ball}, W., {Krivova}, N., {Unruh}, Y., {Haigh}, J., \& {Solanki}, S. 2013,
  Journal of Atmospheric Science, submitted

\bibitem[{{Ball} {et~al.}(2011){Ball}, {Unruh}, {Krivova}, {Solanki}, \&
  {Harder}}]{balletal2011}
{Ball}, W.~T., {Unruh}, Y.~C., {Krivova}, N.~A., {Solanki}, S., \& {Harder},
  J.~W. 2011, \aap, 530, A71

\bibitem[{{Ball} {et~al.}(2012){Ball}, {Unruh}, {Krivova}, {Solanki},
  {Wenzler}, {Mortlock}, \& {Jaffe}}]{balletal2012}
{Ball}, W.~T., {Unruh}, Y.~C., {Krivova}, N.~A., {et~al.} 2012, \aap, 541, A27

\bibitem[{{Berdyugina}(2005)}]{berdyugina2005a}
{Berdyugina}, S.~V. 2005, Living Reviews in Solar Physics, 2, 8

\bibitem[{{B{\"o}hm-Vitense}(2007)}]{stellar_dynamo}
{B{\"o}hm-Vitense}, E. 2007, \apj, 657, 486

\bibitem[{{Borucki} {et~al.}(2010){Borucki}, {Koch}, {Basri}, {Batalha},
  {Brown}, {Caldwell}, {Caldwell}, {Christensen-Dalsgaard}, {Cochran},
  {DeVore}, {Dunham}, {Dupree}, {Gautier}, {Geary}, {Gilliland}, {Gould},
  {Howell}, {Jenkins}, {Kondo}, {Latham}, {Marcy}, {Meibom}, {Kjeldsen},
  {Lissauer}, {Monet}, {Morrison}, {Sasselov}, {Tarter}, {Boss}, {Brownlee},
  {Owen}, {Buzasi}, {Charbonneau}, {Doyle}, {Fortney}, {Ford}, {Holman},
  {Seager}, {Steffen}, {Welsh}, {Rowe}, {Anderson}, {Buchhave}, {Ciardi},
  {Walkowicz}, {Sherry}, {Horch}, {Isaacson}, {Everett}, {Fischer}, {Torres},
  {Johnson}, {Endl}, {MacQueen}, {Bryson}, {Dotson}, {Haas}, {Kolodziejczak},
  {Van Cleve}, {Chandrasekaran}, {Twicken}, {Quintana}, {Clarke}, {Allen},
  {Li}, {Wu}, {Tenenbaum}, {Verner}, {Bruhweiler}, {Barnes}, \&
  {Prsa}}]{KEPLER}
{Borucki}, W.~J., {Koch}, D., {Basri}, G., {et~al.} 2010, Science, 327, 977

\bibitem[{{Deland} \& {Cebula}(2012)}]{delandandcebula2012}
{Deland}, M.~T. \& {Cebula}, R.~P. 2012, Journal of Atmospheric and
  Solar-Terrestrial Physics, 77, 225

\bibitem[{{Domingo} {et~al.}(2009){Domingo}, {Ermolli}, {Fox}, {Fr{\"o}hlich},
  {Haberreiter}, {Krivova}, {Kopp}, {Schmutz}, {Solanki}, {Spruit}, {Unruh}, \&
  {V{\"o}gler}}]{domingoetal2009}
{Domingo}, V., {Ermolli}, I., {Fox}, P., {et~al.} 2009, Space Science Reviews,
  145, 337

\bibitem[{{Dziembowski} {et~al.}(1999){Dziembowski}, {Fiorentini}, {Ricci}, \&
  {Sienkiewicz}}]{dziembowskietal1999}
{Dziembowski}, W.~A., {Fiorentini}, G., {Ricci}, B., \& {Sienkiewicz}, R. 1999,
  \aap, 343, 990

\bibitem[{{Ermolli} {et~al.}(2007){Ermolli}, {Criscuoli}, {Centrone}, {Giorgi},
  \& {Penza}}]{ermollietal2007}
{Ermolli}, I., {Criscuoli}, S., {Centrone}, M., {Giorgi}, F., \& {Penza}, V.
  2007, \aap, 465, 305

\bibitem[{{Ermolli} {et~al.}(2010){Ermolli}, {Criscuoli}, {Uitenbroek},
  {Giorgi}, {Rast}, \& {Solanki}}]{ermollietal2010}
{Ermolli}, I., {Criscuoli}, S., {Uitenbroek}, H., {et~al.} 2010, \aap, 523, A55

\bibitem[{{Ermolli} {et~al.}(2013){Ermolli}, {Matthes}, {Dudok de Wit},
  {Krivova}, {Tourpali}, {Weber}, {Unruh}, {Gray}, {Langematz}, {Pilewskie},
  {Rozanov}, {Schmutz}, {Shapiro}, {Solanki}, \& {Woods}}]{TOSCA2012}
{Ermolli}, I., {Matthes}, K., {Dudok de Wit}, T., {et~al.} 2013, Atmospheric
  Chemistry \& Physics, 13, 3945

\bibitem[{{Fligge} {et~al.}(2000){Fligge}, {Solanki}, \&
  {Unruh}}]{fliggeetal2000}
{Fligge}, M., {Solanki}, S.~K., \& {Unruh}, Y.~C. 2000, \aap, 353, 380

\bibitem[{{Fontenla} {et~al.}(2011){Fontenla}, {Harder}, {Livingston}, {Snow},
  \& {Woods}}]{fontenlaetal2011}
{Fontenla}, J.~M., {Harder}, J., {Livingston}, W., {Snow}, M., \& {Woods}, T.
  2011, Journal of Geophysical Research (Atmospheres), 116, 20108

\bibitem[{{Foukal}(1994)}]{foukal1994}
{Foukal}, P. 1994, Science, 264, 238

\bibitem[{{Foukal}(1998)}]{foukal1998}
{Foukal}, P. 1998, \apj, 500, 958

\bibitem[{{Fr{\"o}hlich}(2006)}]{pmod_comp}
{Fr{\"o}hlich}, C. 2006, \ssr, 125, 53

\bibitem[{{Granzer}(2002)}]{granzer2002}
{Granzer}, T. 2002, Astronomische Nachrichten, 323, 395

\bibitem[{{Gray} {et~al.}(2010){Gray}, {Beer}, {Geller}, {Haigh}, {Lockwood},
  {Matthes}, {Cubasch}, {Fleitmann}, {Harrison}, {Hood}, {Luterbacher},
  {Meehl}, {Shindell}, {van Geel}, \& {White}}]{grayetal2010}
{Gray}, L.~J., {Beer}, J., {Geller}, M., {et~al.} 2010, Reviews of Geophysics,
  48, 4001

\bibitem[{{Haigh}(2007)}]{haigh2007}
{Haigh}, J.~D. 2007, Living Reviews in Solar Physics, 4, 2

\bibitem[{{Hall}(2008)}]{Hall_LR}
{Hall}, J.~C. 2008, Living Reviews in Solar Physics, 5, 2

\bibitem[{{Hall} {et~al.}(2007{\natexlab{a}}){Hall}, {Henry}, \&
  {Lockwood}}]{halletal2007a}
{Hall}, J.~C., {Henry}, G.~W., \& {Lockwood}, G.~W. 2007{\natexlab{a}}, \aj,
  133, 2206

\bibitem[{{Hall} {et~al.}(2009){Hall}, {Henry}, {Lockwood}, {Skiff}, \&
  {Saar}}]{halletal2009}
{Hall}, J.~C., {Henry}, G.~W., {Lockwood}, G.~W., {Skiff}, B.~A., \& {Saar},
  S.~H. 2009, \aj, 138, 312

\bibitem[{{Hall} {et~al.}(2007{\natexlab{b}}){Hall}, {Lockwood}, \&
  {Skiff}}]{halletal2007b}
{Hall}, J.~C., {Lockwood}, G.~W., \& {Skiff}, B.~A. 2007{\natexlab{b}}, \aj,
  133, 862

\bibitem[{{Harder} {et~al.}(2009){Harder}, {Fontenla}, {Pilewskie}, {Richard},
  \& {Woods}}]{harderetal2009}
{Harder}, J.~W., {Fontenla}, J.~M., {Pilewskie}, P., {Richard}, E.~C., \&
  {Woods}, T.~N. 2009, \grl, 36, 7801

\bibitem[{{Hoyt} {et~al.}(1992){Hoyt}, {Kyle}, {Hickey}, \&
  {Maschhoff}}]{hoytetal1992}
{Hoyt}, D.~V., {Kyle}, H.~L., {Hickey}, J.~R., \& {Maschhoff}, R.~H. 1992,
  \jgr, 97, 51

\bibitem[{{Jerzykiewicz} \& {Serkowski}(1966)}]{Lowell_start}
{Jerzykiewicz}, M. \& {Serkowski}, K. 1966, Lowell Observatory Bulletin, 6, 295

\bibitem[{{Judge} {et~al.}(2012){Judge}, {Lockwood}, {Radick}, {Henry},
  {Shapiro}, {Schmutz}, \& {Lindsey}}]{analysis}
{Judge}, P.~G., {Lockwood}, G.~W., {Radick}, R.~R., {et~al.} 2012, \aap, 544,
  A88

\bibitem[{{Keil} {et~al.}(1998){Keil}, {Henry}, \& {Fleck}}]{CAII_SP}
{Keil}, S.~L., {Henry}, T.~W., \& {Fleck}, B. 1998, in Astronomical Society of
  the Pacific Conference Series, Vol. 140, Synoptic Solar Physics, ed. K.~S.
  {Balasubramaniam}, J.~{Harvey}, \& D.~{Rabin}, 301

\bibitem[{{Knaack} {et~al.}(2001){Knaack}, {Fligge}, {Solanki}, \&
  {Unruh}}]{knaacketal2001}
{Knaack}, R., {Fligge}, M., {Solanki}, S.~K., \& {Unruh}, Y.~C. 2001, \aap,
  376, 1080

\bibitem[{{Krivova} \& {Solanki}(2008)}]{krivovasolanki2008}
{Krivova}, N.~A. \& {Solanki}, S.~K. 2008, Journal of Astrophysics and
  Astronomy, 29, 151

\bibitem[{{Krivova} {et~al.}(2003){Krivova}, {Solanki}, {Fligge}, \&
  {Unruh}}]{krivovaetal2003}
{Krivova}, N.~A., {Solanki}, S.~K., {Fligge}, M., \& {Unruh}, Y.~C. 2003, \aap,
  399, L1

\bibitem[{{Krivova} {et~al.}(2011){Krivova}, {Solanki}, \& {Unruh}}]{SATIRE}
{Krivova}, N.~A., {Solanki}, S.~K., \& {Unruh}, Y.~C. 2011, Journal of
  Atmospheric and Solar-Terrestrial Physics, 73, 223

\bibitem[{{Krivova} {et~al.}(2010){Krivova}, {Vieira}, \&
  {Solanki}}]{krivova_rec2010}
{Krivova}, N.~A., {Vieira}, L.~E.~A., \& {Solanki}, S.~K. 2010, Journal of
  Geophysical Research (Space Physics), 115, 12112

\bibitem[{{Lean} {et~al.}(2005){Lean}, {Rottman}, {Harder}, \&
  {Kopp}}]{leanetal2005}
{Lean}, J., {Rottman}, G., {Harder}, J., \& {Kopp}, G. 2005, \solphys, 230, 27

\bibitem[{{Lean} \& {DeLand}(2012)}]{leandeland2012}
{Lean}, J.~L. \& {DeLand}, M.~T. 2012, Journal of Climate, 25, 2555

\bibitem[{{Lockwood} {et~al.}(1992){Lockwood}, {Skiff}, {Baliunas}, \&
  {Radick}}]{lockwoodetal1992}
{Lockwood}, G.~W., {Skiff}, B.~A., {Baliunas}, S.~L., \& {Radick}, R.~R. 1992,
  \nat, 360, 653

\bibitem[{{Lockwood} {et~al.}(2007){Lockwood}, {Skiff}, {Henry}, {Henry},
  {Radick}, {Baliunas}, {Donahue}, \& {Soon}}]{lockwoodetal2007}
{Lockwood}, G.~W., {Skiff}, B.~A., {Henry}, G.~W., {et~al.} 2007, \apjs, 171,
  260

\bibitem[{{Lockwood} {et~al.}(1997){Lockwood}, {Skiff}, \&
  {Radick}}]{lockwoodetal1997}
{Lockwood}, G.~W., {Skiff}, B.~A., \& {Radick}, R.~R. 1997, \apj, 485, 789

\bibitem[{{Noyes} {et~al.}(1984){Noyes}, {Hartmann}, {Baliunas}, {Duncan}, \&
  {Vaughan}}]{noyesetal1984}
{Noyes}, R.~W., {Hartmann}, L.~W., {Baliunas}, S.~L., {Duncan}, D.~K., \&
  {Vaughan}, A.~H. 1984, \apj, 279, 763

\bibitem[{{Oberl{\"a}nder} {et~al.}(2012){Oberl{\"a}nder}, {Langematz},
  {Matthes}, {Kunze}, {Kubin}, {Harder}, {Krivova}, {Solanki}, {Pagaran}, \&
  {Weber}}]{oberlanderetal2012}
{Oberl{\"a}nder}, S., {Langematz}, U., {Matthes}, K., {et~al.} 2012, \grl, 39,
  1801

\bibitem[{{Petit} {et~al.}(2008){Petit}, {Dintrans}, {Solanki}, {Donati},
  {Auri{\`e}re}, {Ligni{\`e}res}, {Morin}, {Paletou}, {Ramirez Velez},
  {Catala}, \& {Fares}}]{petitetal2008}
{Petit}, P., {Dintrans}, B., {Solanki}, S.~K., {et~al.} 2008, \mnras, 388, 80

\bibitem[{{Preminger} {et~al.}(2011){Preminger}, {Chapman}, \&
  {Cookson}}]{premingeretal2011}
{Preminger}, D.~G., {Chapman}, G.~A., \& {Cookson}, A.~M. 2011, \apjl, 739, L45

\bibitem[{{Radick} {et~al.}(1998){Radick}, {Lockwood}, {Skiff}, \&
  {Baliunas}}]{radicketal1998}
{Radick}, R.~R., {Lockwood}, G.~W., {Skiff}, B.~A., \& {Baliunas}, S.~L. 1998,
  \apjs, 118, 239

\bibitem[{{Radick} {et~al.}(1995){Radick}, {Lockwood}, {Skiff}, \&
  {Thompson}}]{radicketal1995}
{Radick}, R.~R., {Lockwood}, G.~W., {Skiff}, B.~A., \& {Thompson}, D.~T. 1995,
  \apj, 452, 332

\bibitem[{{Rauer} {et~al.}(2013){Rauer}, {Catala}, {Aerts}, {Appourchaux},
  {Benz}, {Brandeker}, {Christensen-Dalsgaard}, {Deleuil}, {Gizon}, {Goupil},
  {G{\"u}del}, {Janot-Pacheco}, {Mas-Hesse}, {Pagano}, {Piotto}, {Pollacco},
  {Santos}, {Smith}, {-C.}, {Su{\'a}rez}, {Szab{\'o}}, {Udry}, {Adibekyan},
  {Alibert}, {Almenara}, {Amaro-Seoane}, {Ammler-von Eiff}, {Asplund},
  {Antonello}, {Ball}, {Barnes}, {Baudin}, {Belkacem}, {Bergemann}, {Bihain},
  {Birch}, {Bonfils}, {Boisse}, {Bonomo}, {Borsa}, {Brand{\~a}o}, \&
  {Brocato}}]{PLATO}
{Rauer}, H., {Catala}, C., {Aerts}, C., {et~al.} 2013, ArXiv e-prints

\bibitem[{{Schatten}(1993)}]{schatten1993}
{Schatten}, K.~H. 1993, \jgr, 98, 18907

\bibitem[{{Schrijver} \& {Title}(2001)}]{SchrijverandTitle2001}
{Schrijver}, C.~J. \& {Title}, A.~M. 2001, \apj, 551, 1099

\bibitem[{{Sch{\"u}ssler} {et~al.}(1996){Sch{\"u}ssler}, {Caligari},
  {Ferriz-Mas}, {Solanki}, \& {Stix}}]{polar2}
{Sch{\"u}ssler}, M., {Caligari}, P., {Ferriz-Mas}, A., {Solanki}, S.~K., \&
  {Stix}, M. 1996, \aap, 314, 503

\bibitem[{{Sch{\"u}ssler} \& {Solanki}(1992)}]{polar1}
{Sch{\"u}ssler}, M. \& {Solanki}, S.~K. 1992, \aap, 264, L13

\bibitem[{{Seleznyov} {et~al.}(2011){Seleznyov}, {Solanki}, \&
  {Krivova}}]{seleznyovetal2011}
{Seleznyov}, A.~D., {Solanki}, S.~K., \& {Krivova}, N.~A. 2011, \aap, 532, A108

\bibitem[{{Shapiro} {et~al.}(2013{\natexlab{a}}){Shapiro}, {Schmutz},
  {Cessateur}, \& {Rozanov}}]{shapiroetal2013_stars}
{Shapiro}, A.~I., {Schmutz}, W., {Cessateur}, G., \& {Rozanov}, E.
  2013{\natexlab{a}}, \aap, 552, A114

\bibitem[{{Shapiro} {et~al.}(2011){Shapiro}, {Schmutz}, {Rozanov}, {Schoell},
  {Haberreiter}, {Shapiro}, \& {Nyeki}}]{shapiro_rec}
{Shapiro}, A.~I., {Schmutz}, W., {Rozanov}, E., {et~al.} 2011, \aap, 529, A67

\bibitem[{{Shapiro} {et~al.}(2013{\natexlab{b}}){Shapiro}, {Rozanov},
  {Shapiro}, {Egorova}, {Harder}, {Weber}, {Smith}, {Schmutz}, \&
  {Peter}}]{shapiroetal2013}
{Shapiro}, A.~V., {Rozanov}, E.~V., {Shapiro}, A.~I., {et~al.}
  2013{\natexlab{b}}, Journal of Geophysical Research (Atmospheres), 118, 3781

\bibitem[{{Skumanich} {et~al.}(1984){Skumanich}, {Lean}, {Livingston}, \&
  {White}}]{CAII_CLV}
{Skumanich}, A., {Lean}, J.~L., {Livingston}, W.~C., \& {White}, O.~R. 1984,
  \apj, 282, 776

\bibitem[{{Soderblom} {et~al.}(1991){Soderblom}, {Duncan}, \&
  {Johnson}}]{soderblometal1991}
{Soderblom}, D.~R., {Duncan}, D.~K., \& {Johnson}, D.~R.~H. 1991, \apj, 375,
  722

\bibitem[{{Solanki} {et~al.}(2013){Solanki}, {Krivova}, \& {Haigh}}]{MPS_AA}
{Solanki}, S.~K., {Krivova}, N.~A., \& {Haigh}, J.~D. 2013, \araa, 51, 311

\bibitem[{{Solanki} \& {Unruh}(2013)}]{solankiandunruh2012}
{Solanki}, S.~K. \& {Unruh}, Y.~C. 2013, Astronomische Nachrichten, 334, 145

\bibitem[{{Soubiran} \& {Triaud}(2004)}]{analogs}
{Soubiran}, C. \& {Triaud}, A. 2004, \aap, 418, 1089

\bibitem[{{Thuillier} {et~al.}(2013){Thuillier}, {Melo}, {Lean}, {Krivova},
  {Bolduc}, {Fomichev}, {Charbonneau}, {Shapiro}, {Schmutz}, \&
  {BolsŽe}}]{gerard_comp}
{Thuillier}, G., {Melo}, S. M.~L., {Lean}, J., {et~al.} 2013, Solar Physics,
  submitted

\bibitem[{{Unruh} {et~al.}(2012){Unruh}, {Ball}, \& {Krivova}}]{unruhetal2012}
{Unruh}, Y.~C., {Ball}, W.~T., \& {Krivova}, N.~A. 2012, Surveys in Geophysics,
  33, 475

\bibitem[{{Unruh} {et~al.}(1999){Unruh}, {Solanki}, \& {Fligge}}]{sat_spectra}
{Unruh}, Y.~C., {Solanki}, S.~K., \& {Fligge}, M. 1999, \aap, 345, 635

\bibitem[{{Usoskin}(2008)}]{usoskin2008}
{Usoskin}, I.~G. 2008, Living Reviews in Solar Physics, 5, 3

\bibitem[{{Vieira} {et~al.}(2012){Vieira}, {Norton}, {Dudok de Wit},
  {Kretzschmar}, {Schmidt}, \& {Cheung}}]{luis2012}
{Vieira}, L.~E.~A., {Norton}, A., {Dudok de Wit}, T., {et~al.} 2012, \grl, 39,
  16104

\bibitem[{{Wehrli} {et~al.}(2013){Wehrli}, {Schmutz}, \&
  {Shapiro}}]{wehrlietal2013}
{Wehrli}, C., {Schmutz}, W., \& {Shapiro}, A.~I. 2013, \aap, 556, L3

\bibitem[{{Wenzler} {et~al.}(2006){Wenzler}, {Solanki}, {Krivova}, \&
  {Fr{\"o}hlich}}]{wenzleretl2006}
{Wenzler}, T., {Solanki}, S.~K., {Krivova}, N.~A., \& {Fr{\"o}hlich}, C. 2006,
  \aap, 460, 583

\bibitem[{{White} {et~al.}(1998){White}, {Livingston}, {Keil}, \&
  {Henry}}]{whiteetal1998}
{White}, O.~R., {Livingston}, W.~C., {Keil}, S.~L., \& {Henry}, T.~W. 1998, in
  Astronomical Society of the Pacific Conference Series, Vol. 140, Synoptic
  Solar Physics, ed. K.~S. {Balasubramaniam}, J.~{Harvey}, \& D.~{Rabin}, 293

\bibitem[{{White} {et~al.}(1992){White}, {Skumanich}, {Lean}, {Livingston}, \&
  {Keil}}]{whiteetal1992}
{White}, O.~R., {Skumanich}, A., {Lean}, J., {Livingston}, W.~C., \& {Keil},
  S.~L. 1992, \pasp, 104, 1139

\bibitem[{{Wilson}(1978)}]{wilson1978}
{Wilson}, O.~C. 1978, \apj, 226, 379

\bibitem[{{Woods}(2012)}]{woods2012}
{Woods}, T. 2012, in EGU General Assembly Conference Abstracts, Vol.~14, EGU
  General Assembly Conference Abstracts, ed. A.~{Abbasi} \& N.~{Giesen}, 1520

\end{thebibliography}

\newpage

\begin{appendix}

\section{Deviations in the disc area coverages by active regions}\label{app:dev}
The relationships between disc area coverages and chromospheric activity employed in the present study were established on the basis of solar data  and then extrapolated to higher activity levels. 
While one might expect that the extrapolation works well for stars with activities similar to that of the Sun, the disc area coverages of more active stars may deviate from the values given by Eqs.~(\ref{f_spot})--(\ref{f_fac}). To estimate the impact of such deviations on our results we recalculated  $\log({\rm rms} (b+y)/2) $ and $\Delta[(b+y)/2]/\Delta S$ values plotted in Figs.~\ref{fig:ampl} and \ref{fig:regr}, first assuming that the coverage of the most active stars from the sample of \cite{lockwoodetal2007}  is 50 \% larger then we expect from the extrapolation from the Sun and than that it is  50 \%  smaller than we expect from solar extrapolation.

Namely, we apply following  correction to the spot area disc coverage:
\begin{equation}
A_S'(S)=A_S(S) \cdot (1+\alpha \, \frac{S-S_{\odot}}{S_{\rm max}-S_{\odot}}),
\end{equation}
where $A_S'(S)$ is the new spot disc area coverage, $A_S(S)$  is the spot disc area coverage given by Eq.~(\ref{f_spot}), $\alpha$ is the coefficient which determines the amplitude of the correction, $S_{\odot}$ is mean solar level of chromospheric activity and $S_{\rm max}$ was chosen to be equal to $0.5$, which is the highest mean chromospheric activity  considered in the present study (see the description of the algorithm employed to produce  Figs.~\ref{fig:ampl} and \ref{fig:regr} in Sect.~\ref{sect:pat}).

\begin{figure}
\resizebox{\hsize}{!}{\includegraphics{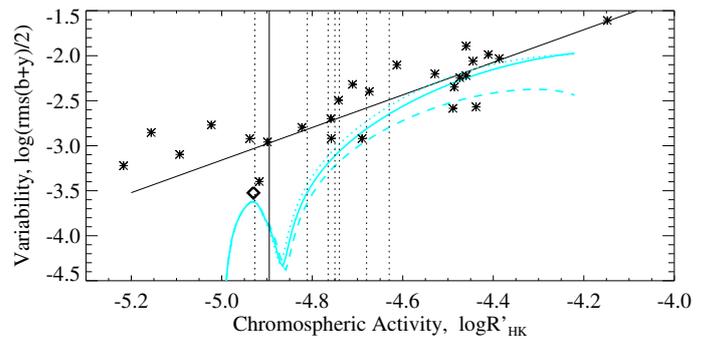}}
\caption{The photometric variability as a function of mean chromospheric activity calculated for $\alpha=0$ (original $A_S$ coverages given by Eq.~\ref{f_spot}, solid curve),  $\alpha=0.5$ (increased $A_S$ coverages, dotted curve), $\alpha=-0.5$ (decreased $A_S$ coverages, dashed curve). The calculations are performed for the homogeneous distribution of active regions.}
\label{fig:ampl_ff}
\end{figure}

\begin{figure}
\resizebox{\hsize}{!}{\includegraphics{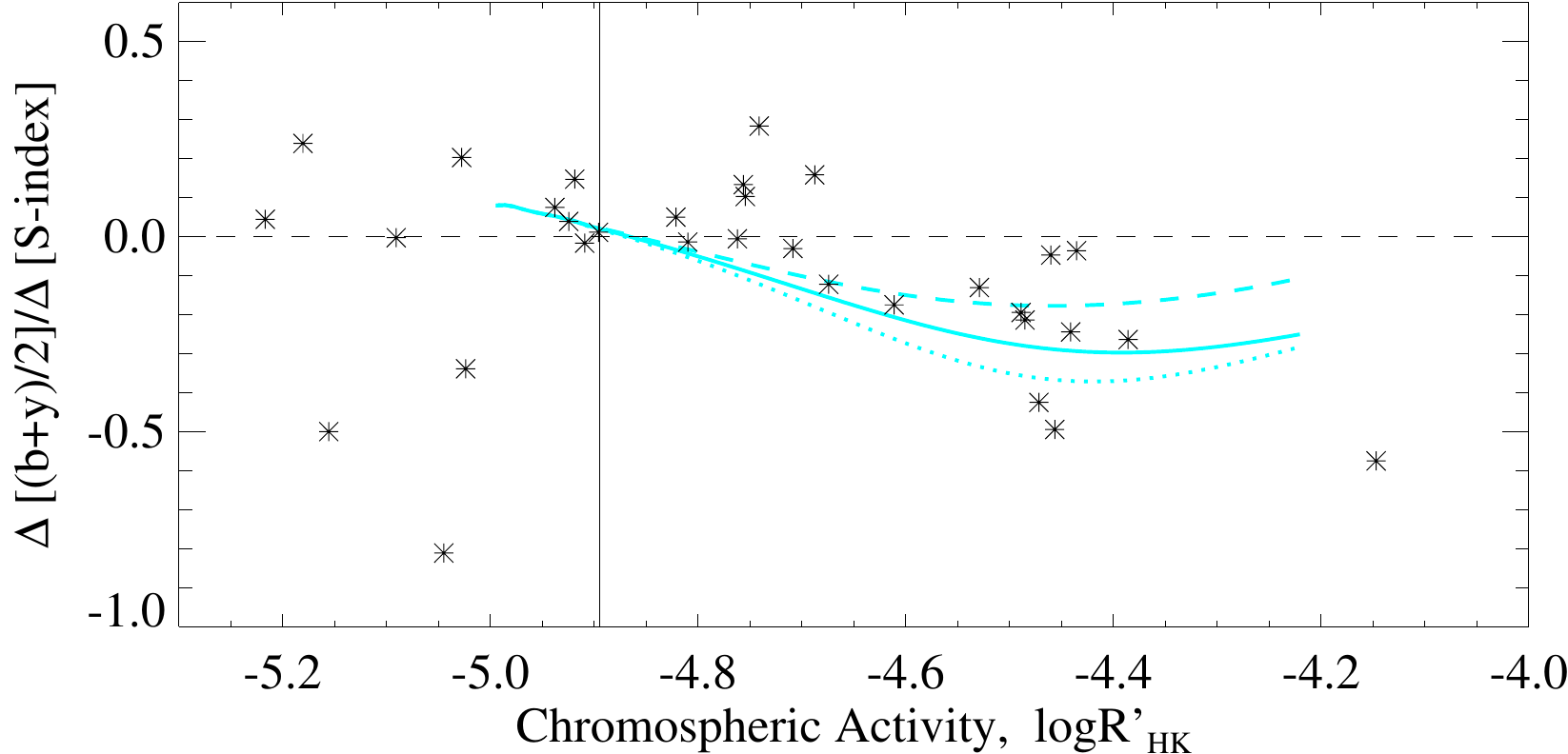}}
\caption{Slope of the regression to the dependence of photometric brightness variation on HK emission variation, plotted vs. mean chromospheric activity ${\rm log R'_{ HK}}$ calculated  for $\alpha=0$ (original $A_S$ coverages given by Eq.~\ref{f_spot}, solid curve),  $\alpha=0.5$ (increased $A_S$ coverages, dotted curve), $\alpha=-0.5$ (decreased $A_S$ coverages, dashed curve).   The calculations are performed for homogeneous distribution of active regions.}
\label{fig:regr_ff}
\end{figure}

The resulting dependences of $\log({\rm rms} (b+y)/2) $ and $\Delta[(b+y)/2]/\Delta S$  on S are plotted in Figs.~\ref{fig:ampl_ff} and \ref{fig:regr_ff} for three values of $\alpha$. One can see that the scatter in the surface coverages may lead to significant deviations in the theoretical curves plotted in Figs.~\ref{fig:ampl_ff} and \ref{fig:regr_ff}.
At the same time the general success of our approach in modeling the stellar data implies that the simple extrapolation of solar disc area coverages is working remarkably well even for stars significantly more active than the Sun.

\section{Comparison with the SATIRE-S results}\label{sect:ratio}
Our model of stellar variability is based on the representation of the facular and spot disc area coverages as functions of the S-index  measured from a vantage point in the stellar equatorial plane (see Eqs.~\ref{f_spot}--\ref{f_fac}). For fixed inclination and distribution of active regions the change of the stellar brightness due to magnetic activity, $\delta (b+y) /2$, is a single-valued function of the observed S-index.

In reality Eqs.~(\ref{f_spot})--(\ref{f_fac}) are only approximate. For every particular observational season the disc area coverages may differ from the values given by these equations, as they define the relation averaged over the longest time interval for which solar data are available (see Sect.~\ref{sect:FF}). For example, a transit of a large spot may cause the stars  deemed facular-dominated by our analysis  to be temporarily spot-dominated.

To estimate the importance of this effect we considered the solar S-index and photometric brightness (i.e. the Str{\"o}mgren $(b+y)$/2 flux) time series. The S-index was calculated from the Sac Peak K-index $K_{\rm SP}$ (see Sect.~\ref{sect:FF}). As there are no long-term solar irradiance measurements equivalent to the Str{\"o}mgren $(b+y)$/2 flux we employed the SATIRE-S spectral irradiance time series \citep[see description in ][and references therein]{balletal2012, ball_climate} convolved with the Str{\"o}mgren $(b+y)$/2 spectral filter profile.
Using these data we calculated the slope of the regression of photometric brightness on the observed S-index, $\Delta[(b+y)/2]/\Delta S$, for 11-year time intervals (the value for the year X is the slope of the regression calculated for the [X-5, X+5] dataset), offset by one year each. 

\begin{figure}
\resizebox{\hsize}{!}{\includegraphics{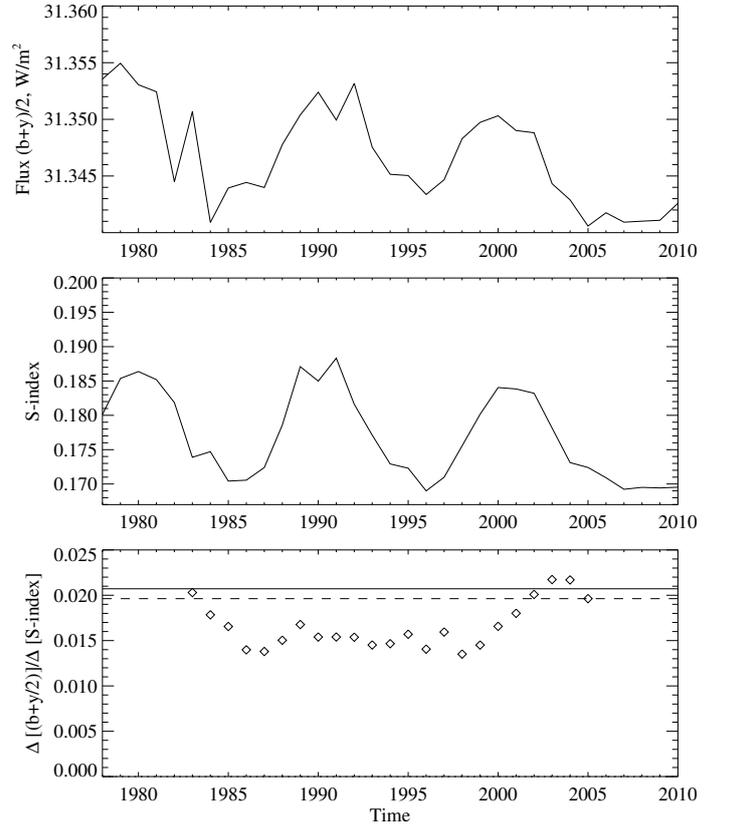}}
\caption{Annual values of the solar spectral flux in the Str{\"o}mgren $(b+y)$/2 filters according to the SATIRE-S model (upper panel) and the S-index calculated from the Sac Peak measurements (middle panel) as well as the slopes of the activity-brightness correlation (lower panel). The solid line in the lower panel represents the slope calculated with the simplified model used in this paper, while the dashed line corresponds to the slope calculated with SATIRE-S data using the entire time series (1978-2010).}
\label{fig:regr_y}
\end{figure}

The S-index and photometric time series and slopes are plotted in Fig.~\ref{fig:regr_y}. One can see that the photometric flux is not a single-valued function of the S-index and the slope of the regression varies with time. For example, the increase of the slope after 2003 might be explained by the decrease of the ratio between spot and facular disc area coverages.

\begin{figure}
\resizebox{\hsize}{!}{\includegraphics{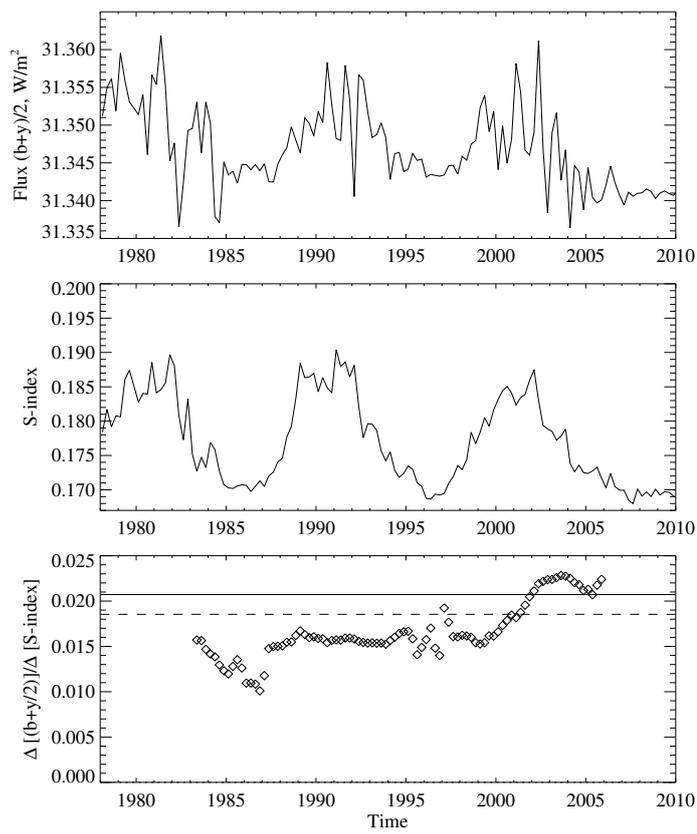}}
\caption{The same as Fig.~\ref{fig:regr_y} but for 3-month averages.}
\label{fig:regr_3m}
\end{figure}

The variations of the slope  may explain some of the scatter in the observed stellar slopes (see Fig.~\ref{fig:regr}). 
The effect gets larger if instead of the annual data we consider 3-month averages (see Fig.~\ref{fig:regr_3m}). One can see that the transit of large spots affects the photometric flux, while leaving the S-index unchanged (as for a star with a solar level of activity the contribution of spots to the S-index is negligibly small). We note that the slopes  calculated with our model using the entire time series (1978-2010)  agree well with the more sophisticated and accurate SATIRE-S calculations.

Figures~\ref{fig:regr_y} and \ref{fig:regr_3m} reveal that according to the SATIRE-S model the variability of the solar Str{\"o}mgren $(b+y)$/2 flux is always faculae-dominated if the Sun is observed for at least 11 years. This does not contradict with its position in the shaded band in Fig.~\ref{fig:regr}  because  if the Sun is observed for a shorter period of time around its activity maximum it can be falsely identified as spot-dominated (see Fig.~\ref{fig:Sun_angles}). 

\Online

\begin{figure*}
\resizebox{\hsize}{!}{\includegraphics{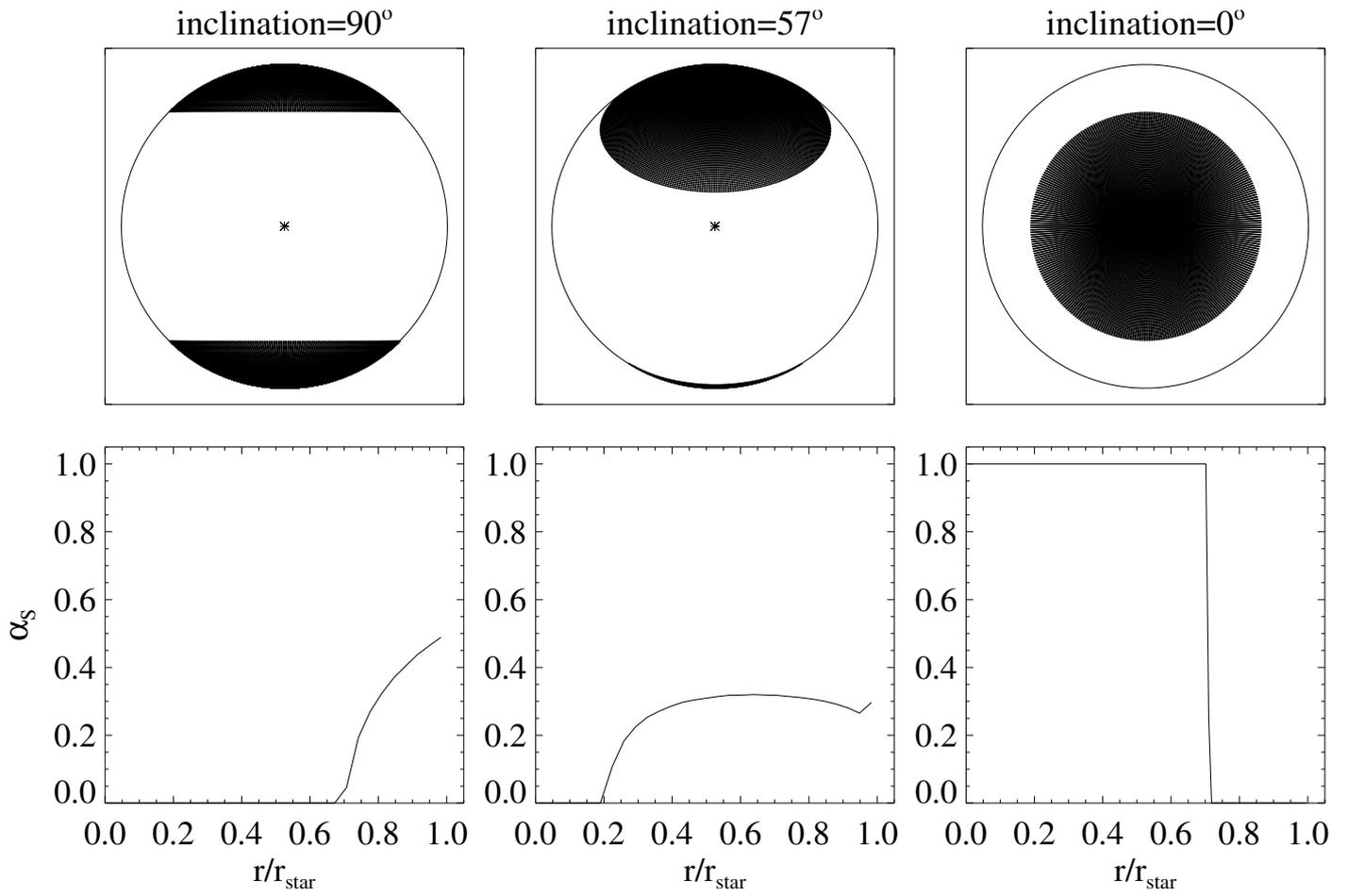}}
\caption{The same as Fig.~\ref{fig:ff_solar} but for the adopted polar distribution of spots (i.e. two caps with latitudes between $\pm 45^\circ$  and $\pm90^\circ$).}
\label{fig:ff_polar}
\end{figure*}

\end{appendix}

\end{document}